\documentclass[fleqn,usenatbib]{mnras}

\usepackage{newtxtext,newtxmath}

\usepackage[T1]{fontenc}

\DeclareRobustCommand{\VAN}[3]{#2}
\let\VANthebibliography\thebibliography
\def\thebibliography{\DeclareRobustCommand{\VAN}[3]{##3}\VANthebibliography}

\usepackage{graphicx}	
\usepackage{amsmath}	
\usepackage{hyperref}
\usepackage{bm}

\newcommand{\hi}{H\textsc{i}}
\newcommand{\althi}{H{\normalfont\textsc{i}}}

\title[HI IM with SKA-Low]{Detecting the \althi\ Power Spectrum in the Post-Reionization Universe with SKA-Low }

\author[Z. Chen et al.]{
Zhaoting Chen,$^{1}$\thanks{E-mail: zhaoting.chen@manchester.ac.uk}
Emma Chapman,$^{2}$
Laura Wolz$^{1}$
and Aishrila Mazumder$^{1}$
\\
$^{1}$Jodrell Bank Centre for Astrophysics, Department of Physics and Astronomy, The University of Manchester, Manchester M13 9PL, UK\\
$^{2}$School of Physics and Astronomy, The University of Nottingham, Nottingham, NG7 2RD, UK\\
}

\pubyear{2022}

\begin{document}
\label{firstpage}
\pagerange{\pageref{firstpage}--\pageref{lastpage}}
\maketitle

\begin{abstract}
We present a survey strategy to detect the neutral hydrogen (\hi) power spectrum at $5<z<6$ using the SKA-Low radio telescope in presence of foregrounds and instrumental effects. 
We simulate observations of the inherently weak HI signal post-reionization with varying levels of noise and contamination with foreground amplitudes equivalent to residuals after sky model subtraction. 
We find that blind signal separation methods on imaged data are required in order to recover the \hi\ signal at large cosmological scales. 
Comparing different methods of foreground cleaning, we find that Gaussian Process Regression (GPR) performs better than Principle Component Analysis (PCA), with the key difference being that GPR uses smooth kernels for the total data covariance. The integration time of one field needs to be larger than $\sim 250$ h to provide large enough signal-to-noise ratio (SNR) to accurately model the data covariance for foreground cleaning. Images within the primary beam field-of-view give measurements of the \hi\ power spectrum at scales $k\sim 0.02\,{\rm Mpc^{-1}}-0.3\,{\rm Mpc^{-1} }$ with SNR $\sim 2-5$ in $\Delta[{\rm log}( k/{\rm Mpc^{-1}})] = 0.25$ bins assuming an integration time of $600$ h. Systematic effects, which introduce small-scale fluctuations across frequency channels, {need to be $\lesssim 5\times 10^{-5}$ to enable unbiased measurements outside the foreground wedge.} Our results provide an important validation towards using the SKA-Low array for measuring the \hi\ power spectrum in the post-reionization Universe.
\end{abstract}




\section{Introduction}

The standard model of cosmology, the $\Lambda$ cold dark matter ($\Lambda$CDM) model, helps us describe and understand the observed Universe. In particular, measurements of the cosmic microwave background (CMB, e.g. \citealt{2020A&A...641A...1P}) and the large scale structure (LSS, e.g. \citealt{2021PhRvD.103h3533A}) can be well fitted by the $\Lambda$CDM model, producing precise, per-cent level constraints on the model parameters. However, as we reach further into the realm of precision cosmology, potential inconsistency between different probes arises in the form of cosmological tensions. Namely, measurements of the Hubble parameter {in} the local Universe using tip of the red-giant branch (TRGB) and Type Ia Supernovae (e.g. \citealt{2022ApJ...934L...7R}) have significant discrepancies $\sim 5\sigma$ with the measurements made using the CMB (e.g. \citealt{2020A&A...641A...6P}). There also exists a tension of $\sim 2.7\sigma$ between the measurements of the amplitude of the dark matter clustering $S_8$ from the CMB and from the LSS (e.g. \citealt{2022PhRvD.105b3514A}).

The disagreements between different cosmological observations highlight the need for understanding the evolutionary history of the Universe. The CMB captures the cosmic structure at the last scattering surface $z\sim1100$ \citep{dodelson2020modern} while the local measurements are made at $z\lesssim 2.0$, missing a large part of the observable Universe in between. One promising approach to fill the gap is neutral hydrogen (\hi) intensity mapping (e.g. \citealt{2004MNRAS.355.1339B, 2008PhRvL.100i1303C, 2008PhRvD..78b3529M,2009MNRAS.397.1926W,2013MNRAS.434.1239B,2017arXiv170909066K}). It uses the emission line of the \hi\ atoms, at the rest wavelength of $\sim 21$ cm, as a tracer of the underlying dark matter distribution. Neutral hydrogen is the most abundant element in the Universe after recombination as predicted by the Big Bang nucleosynthesis \citep{1948PhRv...73..803A, dodelson2020modern}. The formation of dark matter structures, i.e. dark matter halos, attracts baryonic matter to fall into the halos and produces luminous stars and galaxies during the cosmic dawn \citep{2002A&A...382...28S}. The ultra-violet radiation produced by these objects ionized the initially neutral inter-galactic medium (IGM), {a process known as the cosmic reionization \citep{2006PhR...433..181F}. The 21\,cm emission is dominated by the \hi\ inside the IGM during the cosmic reionization}, after which the majority of the remaining \hi\ resides in the dark matter halos \citep{2013MNRAS.430.2427R}. Therefore, the \hi\ signal traces different cosmic structures during different epochs and can be used to probe cosmology across a wide range of redshifts. 

The spectroscopic nature of the 21\,cm line allows the measurement of the matter clustering across the history of structure formation from the cosmic Dark Ages, to the Epoch of Reionization (EoR), and all the way to the low-redshift Universe. {However, the \hi\ signal is inherently weak, and resolving \hi\ sources requires deep integration time even for observing the \hi\ galaxies in the local Universe (e.g. \citealt{2018ApJ...861...49H}).} Without the need to resolve individual sources of the \hi\ emission, intensity mapping {is a technique that maps the 21\,cm emission across a large area of the sky with {relatively coarse angular resolution}, allowing efficient surveys of large cosmological volumes suitable for testing the $\Lambda$CDM model. Ongoing experiments targeting different redshifts include MeerKAT \citep{2016mks..confE..32S}, Canadian Hydrogen Intensity Mapping Experiment (CHIME; \citealt{2022ApJS..261...29C}), Tianlai \citep{2015ApJ...798...40X}, Hydrogen Epoch of Reionization Array (HERA; \citealt{2017PASP..129d5001D}), Low-Frequency Array (LOFAR; \citealt{2017ApJ...838...65P}), Murchison Widefield Array (MWA; \citealt{2013PASA...30....7T}) and more, covering $z\sim 0.0-10.0$. In the future, the Square Kilometre Array Observatory (SKAO) will further enable detections of the neutral hydrogen clustering, {with SKA-Low observing at 50MHz to 350MHz, covering the redshift range from the cosmic Dark Ages $z\sim 27$ down to the post-EoR Universe $z\sim 3.0$ \citep{2015aska.confE...1K}, and SKA-Mid observing at 350MHz to 15.4GHz covering $z\lesssim 3$ \citep{2020PASA...37....7S}.}

The biggest challenge of \hi\ intensity mapping is measuring the signal against the foregrounds which are several orders of magnitude brighter than the \hi. In order to measure the \hi\ signal, extreme accuracy in the instrument calibration is necessary to model the foregrounds \citep{2016MNRAS.461.3135B}. The desired calibration accuracy calls for {a thorough understanding of the sky (e.g. \citealt{2017PASA...34...61T,2018ApJ...869...25M}), the beam (e.g. \citealt{2015ApJ...804...14T,2016ApJ...831..196E})}, and the systematics (e.g. \citealt{2018ApJ...867...15T}). Techniques of foreground mitigation can then be utilised to extract the \hi\ signal. The spectral smoothness of the foregrounds contrasts with the \hi\, which is {discretely structured} in frequency since, for the HI, different frequencies correspond to different redshifts, and therefore different line-of-sight distances. {Fourier transformation along the frequency direction to the delay time space for individual baselines,} a technique called the `delay transform', can thus be used to isolate modes of the power spectrum where the \hi\ dominates \citep{2004ApJ...615....7M,2012ApJ...753...81P,2012ApJ...756..165P}. The region of the wavenumber $k$-space where \hi\ signal can be measured is called the `observation window' whereas the region dominated by the foregrounds is the `foreground wedge' \citep{Datta2010,Morales2012,2014PhRvD..90b3018L}. Measuring the \hi\ power spectrum in the observation window is usually referred to as `foreground avoidance', which is one approach among ongoing efforts of measuring the EoR signal. {Alternatively and/or additionally, Blind signal separation (BSS) techniques can also be applied on the foregrounds, or the residuals of them after sky model subtraction.} These techniques work mostly on the frequency-frequency covariance of the data, such as fast independent component analysis (fastICA, \citealt{2012MNRAS.423.2518C}; \citealt{2014MNRAS.441.3271W}), generalized morphological component analysis (GMCA, \citealt{2013MNRAS.429..165C}), correlated component analysis (CCA; \citealt{2015MNRAS.447.1973B}), gaussian process regression (GPR; \citealt{2018MNRAS.478.3640M}) and more (see \citealt{2019arXiv190912369C} for a review). For \hi\ observations targeting the post-reionization Universe, foreground removal using BSS methods is typically used to recover the \hi\ signal, with transfer function corrections of signal loss (e.g. \citealt{2015ApJ...815...51S}; \citealt{2023MNRAS.523.2453C}).

Using the methods mentioned above, progress has been made at different redshifts towards the detection of the \hi\ power spectrum. For single dish experiments targeting the low-redshift Universe, cross-correlation detections of the \hi\ signal with optical galaxies have been made by the \textit{Green Bank Telescope} \citep{2013ApJ...763L..20M,2013MNRAS.434L..46S,2022MNRAS.510.3495W}, the \textit{Parkes} telescope \citep{2018MNRAS.476.3382A} and the \textit{MeerKAT} telescope \citep{2023MNRAS.518.6262C}. A similar cross-correlation measurement has also been made by the CHIME telescope using stacking \citep{2023ApJ...947...16A}. The first auto-correlation detection has been made by using the \textit{MeerKAT} telescope as a radio interferometer \citep{2023arXiv230111943P}. For experiments targeting EoR, upper limits on the \hi\ power spectrum have been found by the MWA \citep{2016MNRAS.460.4320E,2020MNRAS.493.4711T} and HERA \citep{2022arXiv221004912T} using the delay transform and foreground avoidance, and by LOFAR using map making with GMCA and GPR foreground removal \citep{2017ApJ...838...65P,2020MNRAS.493.1662M}.

In light of the recent progress, in this paper we explore the possibility of measuring the \hi\ power spectrum at $5<z<6$ using SKA-Low. While this redshift range is within the frequency coverage of the instrument, it has been largely neglected since it is not in the interests of the primary goal of \hi\ science for SKA-Low, which mainly focuses on the EoR \citep{2015aska.confE...1K}. Despite probing different physics, observations of the post-reionization Universe can benefit significantly from the wide frequency range of the SKA-Low telescope, as the deep observations of the EoR fields will provide accurate modelling of the radio continuum and the instrument. {Furthermore, it has been suggested that the Universe may still be partially ionized at $z\sim 5.5$ \citep{2022MNRAS.514...55B}, in contrast with conventional constraints on the end of reionization to be at $z\sim 6$ (e.g. \citealt{2006AJ....131.1203F}). Using the \hi\ power spectrum at $5<z<6$ provides a unique method of constraining the end of reionization.} However, measuring the \hi\ clustering at these redshifts has its own challenges. The \hi\ signal at the quasi-linear scales probed at $5<z<6$ will be lower than the signal at the EoR. Meanwhile, the low-frequency band contains more foreground contamination than the L-band typically used for intensity mapping at lower redshifts. It is important to quantify the signal and foreground level at these frequencies as well as the instrument effects, to verify if these redshifts can be used for cosmology.

In this paper, we present an end-to-end pipeline including simulations of the sky signals and the interferometric observations, the foreground mitigation, and the power spectrum estimation to provide a proof-of-concept study for measuring the \hi\ power spectrum at $5<z<6$ using SKA-Low. Using the simulation pipeline with different settings, we explore different levels of foreground residual and noise level to find the requirements on integration time and foreground modelling needed. Methods for residual foreground removal are investigated focusing on the comparison between Principle Component Analysis (PCA) and GPR, with quantitative investigations into the differences in the performance of these two methods. We present our forecasts for future SKA-Low surveys on the power spectrum measurements. Impacts of systematics are also briefly discussed to provide an estimation of the requirements on levels of the systematics.

The paper is organised as follows: The simulation of the sky signal is described in Section \ref{sec:skysim}. Simulations of the interferometric observations to get the images and subsequent power spectrum estimation from the images are discussed in Section \ref{sec:vissim}. The presence and the structure of the foreground wedge, with foreground mitigation methods applied, are quantified in Section \ref{sec:fgwedge}. The robustness of the foreground mitigation methods is tested in the presence of thermal noise and systematic effects in Section \ref{sec:forecast}. We present the concluding remarks in Section \ref{sec:conclusion}. Throughout this paper, we assume the $\Lambda$CDM cosmology from \cite{2020A&A...641A...6P}.

\section{Simulations of the Radio Sky}
\label{sec:skysim}

\begin{figure}
    \centering
	\includegraphics[width=0.9\columnwidth]{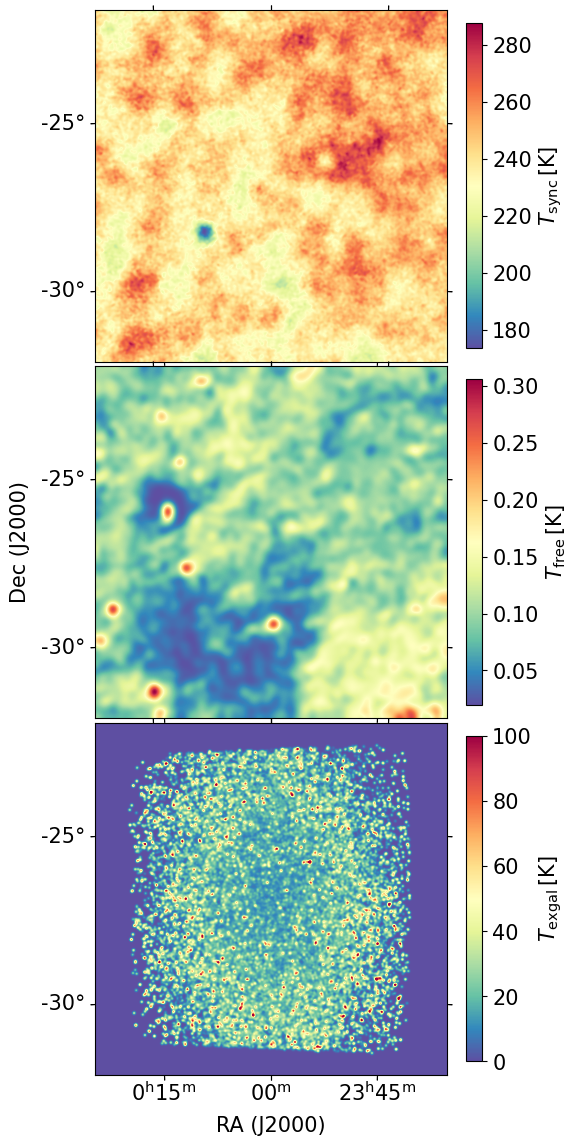}
    \caption{The input sky simulations of different foreground components at 220MHz as described in Section \ref{sec:skysim}. The simulation of the synchrotron radiation is shown in the top panel. The simulation of the free-free emission is shown in the middle panel. The simulation of the extragalactic radio sources are shown in the bottom panel. The pixel size of the figure is $(21\,{\rm arcsecond})^2$ and the total size of the signal simulation is $(10.5\,{\rm deg})^2$. Note that the extragalactic signal shown in the bottom panel is simply for illustration with the sources plotted as point sources. Values larger than 100K are masked for better presentation. When simulating the observations, the radio sources are directly put in as a source catalogue instead of a map, as described in Section \ref{subsec:exgal}.}
    \label{fig:fgsim}
\end{figure}

In this section, we outline the simulations of the sky signal which consist of the \hi\ signal and the foregrounds at $5<z<6$, corresponding to $\sim 200-240$MHz. {The SKA-Low instrument is designed to have a maximum channel resolution of $5.4$kHz \citep{2019arXiv191212699B}. Since we are only interested in the \hi\ intensity mapping which uses large voxels to map the distribution of the \hi\ emission,} we reduce the simulated data volume by assuming the redshift bin is covered by 66 frequency channels with a channel bandwidth of 510kHz. {The coarser frequency resolution of 510kHz corresponds to $k_\parallel \sim 0.4\,{\rm Mpc^{-1}}$. While increasing the frequency resolution gives access to higher $k_\parallel$ where the foregrounds are weaker, the small scales beyond BAO wiggles are difficult to model for cosmological inferences. We leave simulations with the full frequency resolution for future work.} 

The primary beam field-of-view (FoV) for SKA-Low at these frequencies is $\sim 3$ degrees \citep{2019arXiv191212699B}. {We simulate $(10.5 {\rm deg})^2$ sky areas around the pointing centre for all the components of the sky signal. While the sky area only extends to the -20dB (1\%) sidelobes of the primary beam, we find that there is no sharp features in the cylindrical power spectrum from simulated foreground residuals (see Appendix \ref{sec:caveat}). As discussed later in Section \ref{sec:vissim}, we perform the power spectrum estimation using only the centre $(1.5\,{\rm deg})^2$ and therefore the $(10.5 {\rm deg})^2$ sky area is sufficient.} The pointing centre is at the EoR0 field \citep{2021PASA...38...57L} at RA=0h, Dec=-27deg. The methods for generating the components are described as follows.

\subsection{Diffuse Galactic radiation}
The diffuse Galactic radiation at these scales is dominated by the synchrotron radiation. We use the all-sky `Haslam map' of synchrotron radiation at 408MHz \citep{1981A&A...100..209H,1982A&AS...47....1H} with the updated version described in \cite{2015MNRAS.451.4311R}. The map is then extrapolated to the frequencies of interest using the Global Sky Model \citep{2017MNRAS.464.3486Z} at 1.4GHz and 2.3GHz to calculate the spectral indices of the map pixels. The curvature of the spectral indices (see e.g. \citealt{2022MNRAS.509.4923I}) is neglected for simplicity. 

The pixel size of the input Haslam map is $(1.72\,{\rm arcmin})^2$, corresponding to \textsc{HEALPIX} \citep{2005ApJ...622..759G,Zonca2019} NSIDE=2048. An image of $(10.5 \, {\rm deg})^2$ around the pointing centre is created with a pixel size of 21 arcsec. The image is then Gaussian smoothed with a resolution of 1.75 arcmin. The input synchrotron radiation at the central frequency of our simulation 220MHz is shown in Fig. \ref{fig:fgsim}.

Free-free emission from the Galactic electrons also contributes to the diffuse Galactic radiation. Following \cite{2020MNRAS.496.1232L}, we use \textsc{fg21sim} \footnote{\url{https://github.com/ChenxiSSS/FG21SimPlus}} to simulate the Galactic free-free emission. It is based on the H$\alpha$ intensity map in \cite{2003ApJS..146..407F}. The free-free emission in the frequency range of our interest is several orders of magnitude smaller than the synchrotron as shown in Fig. \ref{fig:fgsim}. 

As discussed later in Section \ref{subsec:skytoim}, we make image cubes of the observations to perform residual foreground removal and power spectrum estimation. In interferometric observations, during the calibration and imaging process, the diffuse emission is largely subtracted and no visible structure is left in the image cube (see e.g. \citealt{2022MNRAS.512.2697R}). {
Therefore, in our work, we assume the majority of diffuse emission has been removed and model the diffuse foreground residual amplitude as 0.1\% of the original emission of our simulation.
Although this approach will require accurate modelling of the sky signal, it is fully within the power of SKA-Low. Note that while we are only simulating 66 frequency channels from 200 to 240 MHz, a much wider frequency range, from 50MHz to 350 MHz, will be utilised in future SKA-Low observations to provide accurately modelling of the continuum emission. 
As we show later in Section \ref{subsec:skytoim}, the output foreground image cube fluctuates on the scale of $\sim 2$mJy per point spread function (PSF), corresponding to the overall fluctuation of roughly $80$mJy, consistent with the flux density level of residual image cubes from existing EoR observations (see e.g. fig. 2 of \citealt{2020MNRAS.493.1662M}). Thus, the assumption for the level of foreground residual is representative for SKA-Low. It is beyond the interest of this preliminary work to simulate the entire frequency range and produce the sky model for visibility subtraction.} 

Note that there are other sources of foregrounds that are of Galactic origins, such as supernovae remnants \citep{2015aska.confE..64W}. Since the dominant component of the foregrounds is the synchrotron, we expect that the Galactic foreground simulated in our work is enough to capture the amplitude and the structure of the diffuse emission and leave other components of the Galactic foregrounds for future study.

\subsection{Extragalactic radio sources}
\label{subsec:exgal}
Apart from the Galactic diffuse emission, extragalactic radio sources also contribute to the overall foreground emission. While the Galactic foregrounds are mostly diffuse, the extragalactic foregrounds are typically individual sources of finite size. Understanding the properties of the radio galaxies is a major scientific goal for radio surveys. For example, both continuum and \hi\ science results have been produced using the same fields of the MIGHTEE survey \citep{2022MNRAS.509.2150H,2022ApJ...935L..13S}; Observations of EoR0 field from the MWA are used to produce both the upper limits on the reionization power spectrum and the source catalogue \citep{2016ApJ...833..102B,2020MNRAS.493.4711T,2021PASA...38...57L}.

For future observations using SKA-Low, we expect a good understanding of the radio sources in the fields which will be iteratively improved as the observations themselves will further help build more complete catalogues. Here we use the source catalogue from the LOFAR Two-meter Sky Survey observations of the ELAIS-N1 (EN1) field \citep{2021A&A...648A...2S} and rotate the centre of the field to our pointing centre as shown in Fig. \ref{fig:fgsim}. The EN1 catalogue covers slightly less than the $(10.5\,{\rm deg})^2$ sky area used for simulating the diffuse foregrounds. As discussed later in Section \ref{sec:vissim}, we only image the central $(1.5\,{\rm deg})^2$ fields so the smaller input sky area for the radio sources has negligible impacts on the intensity of the foreground emission in our image cubes. In real observations, the bright sources in the beam sidelobe pose challenges to the data calibration which we do not consider in this work. These issues can be mitigated by techniques such as secondary and direction-dependent calibrations (see e.g. \citealt{2017ApJ...838...65P,2018MNRAS.478.3640M,2022MNRAS.509.2150H}).

In the source catalogue, we impose a flux density cut of 10mJy assuming all sources above this flux density can be perfectly peeled. The 10mJy limit is fairly conservative and can be set lower given the high sensitivity of SKA-Low. For example, using 12 nights of LOFAR-EoR data observing the North Celestial Pole (NCP), \cite{2020MNRAS.493.1662M} produced source-subtracted images with fluctuations at 50mJy level. The source model of the NCP field has also been built iteratively over the years down to sources with flux density down to $\sim 3$mJy \citep{2013A&A...550A.136Y}. The depth of the sky model for the EoR0 field simulated in this work can also be expected to reach mJy level. Furthermore, we expect the sources below this flux density to be modelled with 90\% accuracy. This is again a conservative estimate, as relatively short observations of only 13 h used in \cite{2017ApJ...838...65P} reports $\sim 5\%$ error in recovering the flux density of a known bright source. As we discuss later, we focus on deep observations with $\geq 300$ h of observation and therefore it is expected that {the flux of the sources} around 1mJy can be accurately modelled with below 10\% errors. {We assume no position errors for the sky modelling}.

\subsection{The \althi\ signal}
\label{subsec:hisim}
\hi\ resides mostly inside the dark matter halos after the EoR at $z\lesssim 6$. The collapse of the cold gas leads to star formation, creating strong correlations between the star formation rate and the molecular ($\rm H_2$) gas content of the galaxies \citep{2008AJ....136.2782L}. Therefore, the clustering of \hi\ can be related to the star forming properties of the galaxies and can be used to constrain the galaxy astrophysics (e.g. \citealt{2016MNRAS.458.3399W,2021MNRAS.502.5259C}). At higher redshifts beyond cosmic noon $z>2$, the fraction of \hi\ within galaxies start to drop \citep{2018ApJ...866..135V} and the distribution of the \hi\ tilts more towards the massive halos \citep{2020MNRAS.493.5434S}. Due to the lack of direct observations on these \hi\ emission sources at higher redshifts, the properties of the \hi\ within halos are not well understood, which can be dramatically improved by future \hi\ intensity mapping experiments.

The large sky area of $(10.5\,{\rm deg})^2$, and the $5<z<6$ redshift bin, result in a light cone of $\sim 1500$Mpc in the transverse direction and $\sim 500$Mpc in the line-of-sight (los) direction. For our purposes of exploring the detectability of the signals, instead of using a full hydrodynamical simulation, we use semi-analytical simulations based on dark matter simulations and \hi\ Halo Occupation Distribution (HOD, \citealt{2002PhR...372....1C}). It allows us to efficiently simulate the large volume required. The detailed steps of our \hi\ simulation are as follows:
\begin{itemize}
    \item 
    Assuming the Planck18 cosmology \citep{2020A&A...641A...6P}, we use \textsc{pinocchio}\footnote{\url{https://github.com/pigimonaco/Pinocchio}} \citep{2002MNRAS.331..587M,2013MNRAS.433.2389M} to simulate nine boxes of dark matter distributions, each with a volume of $(620\,{\rm Mpc})^3$. The total volume of $9 \times (620\,{\rm Mpc})^3$ is to ensure that the lightcone falls well within the simulated volume, avoiding edge effects. The total volume is divided into 9 sub-boxes to avoid computational difficulties.
    \item
    Each sub-box has 1850 grid points per side, resulting in a mass resolution of $\sim 3.25 \times 10^9\,{\rm M_\odot /h}$. Note that this mass resolution is likely not enough to resolve all the \hi-rich halos (see e.g. \citealt{2018ApJ...866..135V}). However, it is enough to capture the bias of the \hi\ clustering which is sufficient for our purposes.
    \item
    Each sub-box is simulated across the $5<z<6$ redshift bin with a snapshot taken at each observing frequency channel, equalling a total of 66 snapshots (see Section \ref{sec:vissim} for specifications of the observations). The halo positions relative to the centre of the box in comoving space, the velocities, and the mass of the haloes are taken.
    \item
    The 9 sub-boxes are then put together onto 3x3 grids with the centres of the boxes re-positioned. We take the observer to be at (0,0,0) and the centre of the 5th box is at (0,0,$\rm X_{\rm cen}$) where $\rm X_{\rm cen}$  is the comoving distance at the centre of the $5<z<6$ redshift bin. The halos are re-positioned accordingly.
    \item
    The peculiar velocities of the halos are calculated given the 3D halo velocities and the position vectors. The halo positions are modified to redshift space according to the Kaiser effect \citep{1987MNRAS.227....1K}.
    \item
    Each halo is assigned an \hi\ mass according to the \hi\ HOD of the IllustrisTNG simulation in \cite{2018ApJ...866..135V}. The \hi\ HOD follows $M_{\rm \hi} = M_0 (M_{\rm h}/M_{\rm min})^\alpha{\rm exp}(-(M_{\rm min}/M_{\rm h})^{0.35})$ with $M_{\rm h}$ the halo mass. We adopt the parameter values at $z=5$, with $M_0 = 1.9\times 10^9\,{h^{-1}M_\odot}$, $M_{\rm min} =2.0\times 10^{10}\,{h^{-1}M_\odot}$ and $\alpha = 0.74$. All HI masses are put into the halo centres, since we are only interested in large scales $k < 0.5\, {\rm Mpc^{-1}}$ and hence the halos are unresolved. {The \hi\ masses are then multiplied by} a constant factor so that at each redshift the HI mass density, $\Omega_{\rm \hi}$, equals to $10^{-3}$. This is consistent with the observation of \cite{2015MNRAS.452..217C} and ensures that the clustering amplitude is realistic.
    \item
    The distances between the halos and the observer are calculated. For snapshot $i$ corresponding to frequency channel $i$, the line-of-sight comoving distance range $[{\rm X_{\rm min}^{ i},X_{\rm max}^{ i}}]$ is calculated according to the channel bandwidth and central frequency. Only halos in the distance range are selected.
    \item
    A rotational matrix along $y$ axis to rotate the $x$-$z$ plane is applied to the halos so that the centre of the simulation corresponds to the pointing centre RA=0h and Dec=-27deg. The halo positions are converted to angular coordinates. The \hi\ mass is converted to the flux density assuming that the flux is distributed as a step function across the frequency channel. This is a reasonable assumption given that the velocity resolution of SKA-Low at $5<z<6$ is not high enough to resolve the emission profiles of the \hi\ sources.
\end{itemize}

\begin{figure}
    \centering
	\includegraphics[width=\columnwidth]{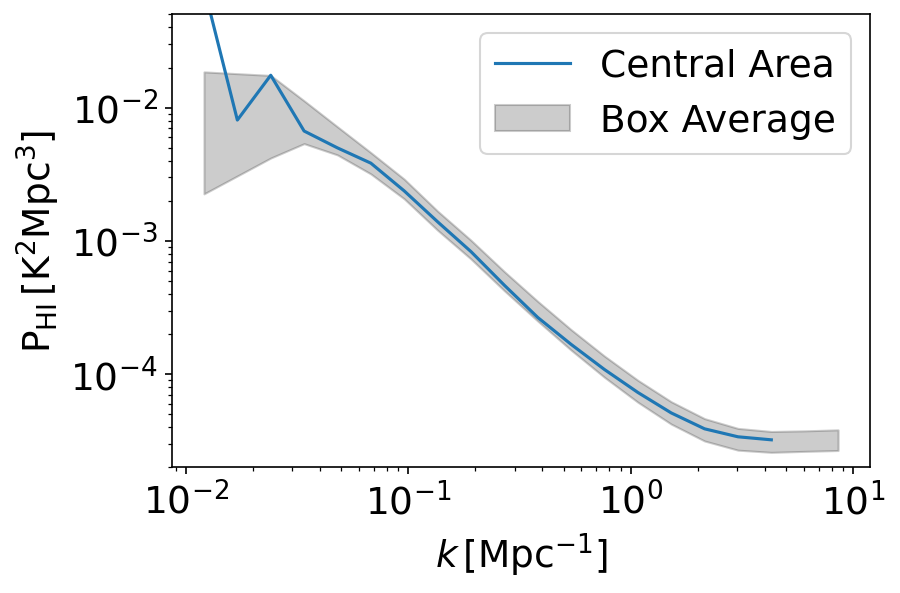}
    \includegraphics[width=\columnwidth]{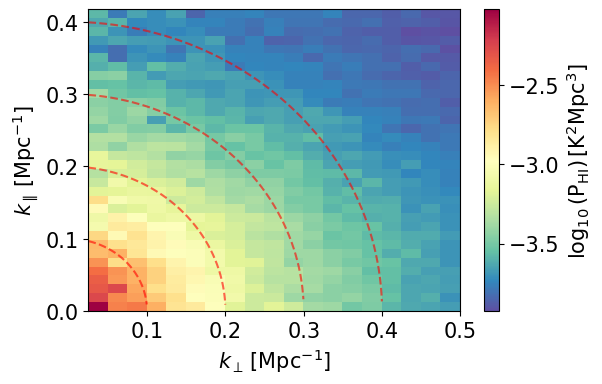}
    \caption{The brightness temperature power spectrum of the \hi\ simulation described in Section \ref{subsec:hisim}. In the top panel the blue solid line shows the 1D \hi\ power spectra for the central area of $\rm (1.5\,deg)^2$ in the simulated lightcone. The shaded area shows the one standard deviation range of the input \hi\ power spectrum where the standard deviation is calculated from all the snapshots of all the sub-boxes. The bottom panel shows the cylindrical power spectrum of the central area of $\rm (1.5\,deg)^2$ in the simulated lightcone. The red dashed line denotes the $k=\{0.1,0.2,0.3,0.4 \}{\rm Mpc^{-1}}$ contours for reference. The \hi\ power spectrum of the central area agrees tightly with the \hi\ power spectrum of the entire box, and is largely isotropic.}
    \label{fig:hisim}
\end{figure}

To validate our \hi\ simulation, we compute the \hi\ power spectra for the 9 sub-boxes, and compare to the central $\rm (1.5\, deg)^2$ area of the light cone which we will use for imaging later. The resulting average \hi\ power spectra for the boxes and for the central input image is shown in Fig. \ref{fig:hisim}. 

{We emphasize that the variance of the \hi\ signal, shown as the shaded area in Fig. \ref{fig:hisim}, is underestimated. This is due to the fact that we assume a deterministic relation between the \hi\ and halo mass, ignoring the scatter of the relation (see e.g. Fig.4 of \citealt{2018ApJ...866..135V}). The scatter comes from the assembly bias of halos, which can be introduced by the inhomogenous reionization history (e.g. \citealt{2022arXiv221002385L}). In our case of investigating the detectability in thermal noise dominated case, this effect is negligible and we leave more realistic simulations for future work.}

Note that, the $(\rm 1.5 \, deg)^2$ image size corresponds to a maximum length scale equivelant to $k\sim 0.03 \,{\rm Mpc^{-1}}$. Scales larger than this can not be probed by the image, as one can see from the top panel of Fig. \ref{fig:hisim}. At smaller scales, $k>1\,{\rm Mpc^{-1}}$, the \hi\ power spectrum hits the shot-noise plateau. This is not accurate and the actual shot noise should be much lower. In our simulation, the \hi\ is directly put as point sources in the halo centres, so that the number density of \hi\ sources is underestimated (see \citealt{2020MNRAS.493.5434S} for a discussion of this). The actual shot noise should be much lower and requires more in-depth modeling of the \hi\ halo model \citep{2019MNRAS.484.1007W,2021MNRAS.502.5259C}. As we will discuss in Section \ref{sec:vissim}, the minimum $k$ scale probed in our simulation is $k\sim 0.3\,{\rm Mpc^{-1}}$ and therefore we are not affected by this insufficient modelling. The cylindrical power spectrum shown in the bottom panel of Fig. \ref{fig:hisim} indicates that the \hi\ power spectrum from our simulation gives the correct isotropic features, and therefore can be reliably used to study the detectability of the \hi\ power spectrum in the presence of the foreground wedge.

\section{Simulations of Observations}
\label{sec:vissim}
In this section, we describe the simulation of the SKA-Low interferometer to observe the input sky signal discussed in Section \ref{sec:skysim}, the imaging routine to produce the image cube within the primary beam FoV, and the power spectrum estimation.

\subsection{From sky signal to image product}
\label{subsec:skytoim}
The SKA-Low array will consist of 131,072 log-period dipole antennas within 512 stations covering the southern sky from 50 to 350 MHz. Since the specific station layout and specifications are not finalised, we use the v3 station layout \citep{2020arXiv200312744D} assuming a frequency channel bandwidth of 510kHz. We only take the central area with 296 stations with a maximum baseline length of 3.15km. The longest baselines are not of {cosmological interest} and are thus neglected to reduce data volume. The frequency range we simulate is from 202.56MHz to 235.76MHz, covering redshift 5 to 6 with 66 frequency channels. The station layout is shown at the left panel of Fig \ref{fig:array}.

The visibility data are simulated to represent one night of observation at the EoR0 field. We assume a total integration of 12 h with a time-resolution of 180 s in one tracking. The resulting u-v coverage of the baselines is shown in the right panel of Fig \ref{fig:array}. {The u-v coverage shown is dense within $\rm |u|<1000m$ (which corresponds to the physical scale $k\lesssim 0.5\,{\rm Mpc^{-1}}$). Choosing the u-v grid length to correspond to our image size, we find no loss of u-v grid sampling, justifying the usage of a relatively coarse time resolution.} 

\begin{figure}
    \centering
	\includegraphics[width=\columnwidth]{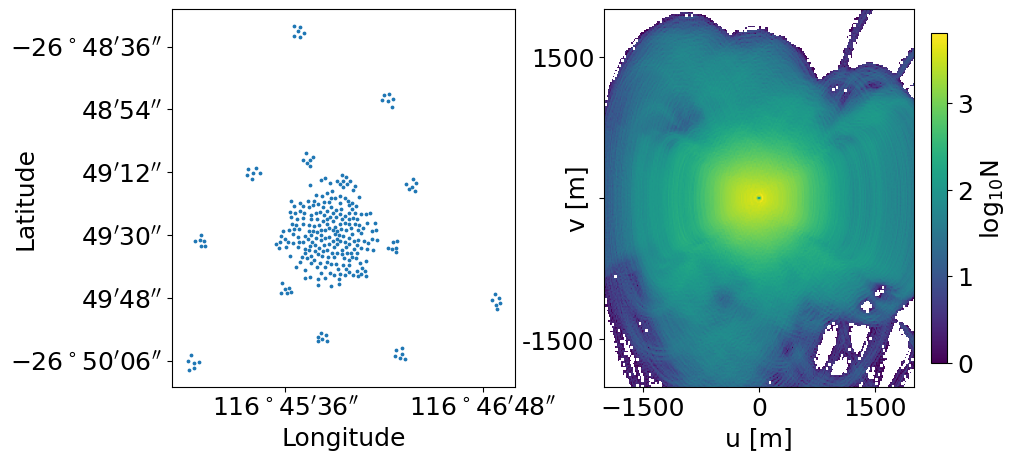}
    \caption{Left panel: The station layout used in our simulation. Each dot denotes one station. Right panel: The u-v distribution of the simulation for a 12-h tracking of the EoR0 field with a time resolution of 180 seconds. The colors denote the number of instantaneous baselines in one u-v grid. Each u-v grid has a size of $\rm 16\times 16\, m^2$.}
    \label{fig:array}
\end{figure}

\begin{figure}
    \centering
	\includegraphics[width=0.95\columnwidth]{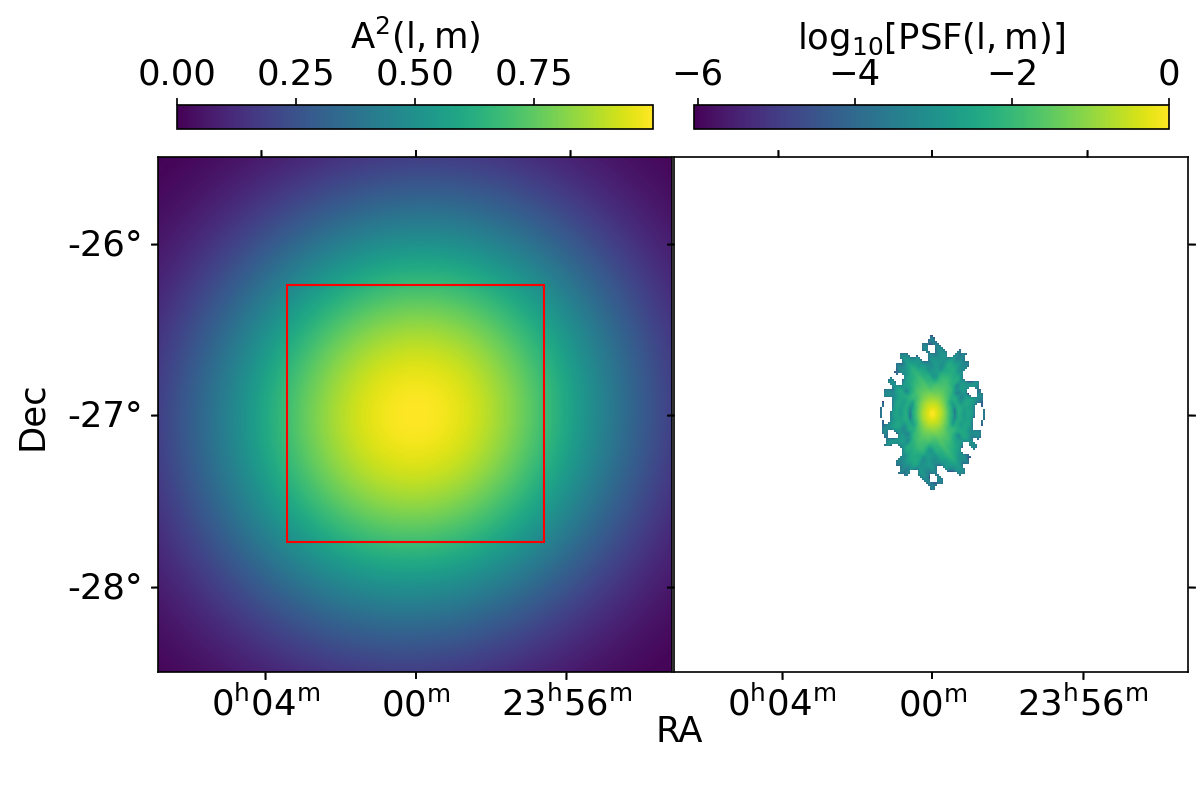}
    \caption{Left panel: The power-square beam $\rm A^2(l,m)$ around the pointing centre in our simulation. The primary beam is averaged across all stations. The red square shows the $\rm (1.5\,deg)^2$ area within which the image cube is produced. Right panel: The point spread function (PSF) corresponding to the u-v coverage of our simulation using natural weighting. Pixels with values $\approx 0$ are left blank. Both figures have a size of $\rm (3\,deg)^2$ with $512\times 512$ pixels. Note that both the primary beam and the PSF are frequency-dependent and we show the values at central frequency 220MHz here for presentation.}
    \label{fig:beam}
\end{figure}

Following the observational specifications discussed above, we use the \textsc{OSKAR}\footnote{\url{https://github.com/OxfordSKA/OSKAR}} package \citep{5613289} to generate the visibility data. \textsc{OSKAR} takes in the telescope specifications, sky model and observation strategy to simulate the primary beam, the u-v coverage and the visibility data. It can also be used to generate dirty images, which we use to produce the image cube. The sky area for the imaging output is determined by the primary beam size. In the calculation of the power spectrum, the primary beam attenuation is squared since the power spectrum is the Fourier density field squared (see e.g. \citealt{2014ApJ...788..106P}). To image within the primary beam field-of-view, we take the limit where the power-square beam attenuation reaches $\sim 0.5$. The primary beam is largely Gaussian near the pointing centre as shown in Fig. \ref{fig:beam}, resulting in power-square beam having half the full-width-half-maximum (FWHM) comparing to the actual beam. The image size is accordingly set to be $\rm (1.5\,deg)^2$ and we choose the pixel size to be $\rm (0.45\, arcmin)^2$ with $200\times200$ grids. We apply the W-projection algorithm \citep{2008ISTSP...2..647C} with natural weighting to the baselines to produce the image cube. The power-square beam and the synthesized beam (PSF) are shown in Fig \ref{fig:beam}. The point spread function in Fourier space has a FWHM of $k\sim 0.3\,{\rm Mpc^{-1}}$. 

Gaussian random noise are added to the visibility data to simulate the thermal noise. The amplitude of the thermal noise is determined by the radiometer equation \citep{2013tra..book.....W}
\begin{equation}
    \sigma_{\rm N} = \frac{2{\rm k_{B}}T_{\rm sys}}{A_{\rm e}\sqrt{\delta f \delta t}},
\label{eq:sigman}
\end{equation}
where $\rm k_B$ is the Boltzmann constant, $T_{\rm sys}$ is the system temperature, $A_{\rm e}$ is the effective collecting area, $\delta f$ is the frequency channel bandwidth, $\delta t$ is the time resolution. We follow \cite{2019arXiv191212699B} and set the natural sensitivity $A_{\rm e}/T_{\rm sys} = 1.235 \,{\rm m^2K^{-1}}$ to generate random complex Gaussian on every baseline. The images at the central frequency for the foregrounds, the \hi, and the thermal noise are shown in Fig. \ref{fig:image}. All images are dirty images with no cleaning routine applied. {Throughout this paper, we use `Jy/PSF' and `kelvin/PSF' units for the images before deconvolution with the PSF. The `PSF' refers to the integrated PSF area in steradian $\int {\rm d}l{\rm d}m\,{\rm PSF}(l,m)$. `Jy/PSF' is more commonly referred to as `Jy/beam'. We use `Jy/PSF' to avoid confusion with the primary beam.} 

In Section \ref{subsec:tnfg} when we discuss residual foreground removal, the thermal noise is rescaled by a factor of $\sqrt{t_{\rm sim}/{t_{\rm int}}}$, where $t_{\rm sim}= 12$ h is the observation time for the simulated one tracking and $t_{\rm int}$ is the total integration time set to 360, 480 and 600 h for different scenarios. The rescaling mimics coherent averaging of the visibility data over multiple nights. {The thermal noise power spectrum is $\sim 4$ orders of magnitude larger than the \hi\ power spectrum as we show in Fig. \ref{fig:bias} in Appendix \ref{sec:est}.}

\begin{figure}
    \centering
	\includegraphics[width=0.99\columnwidth]{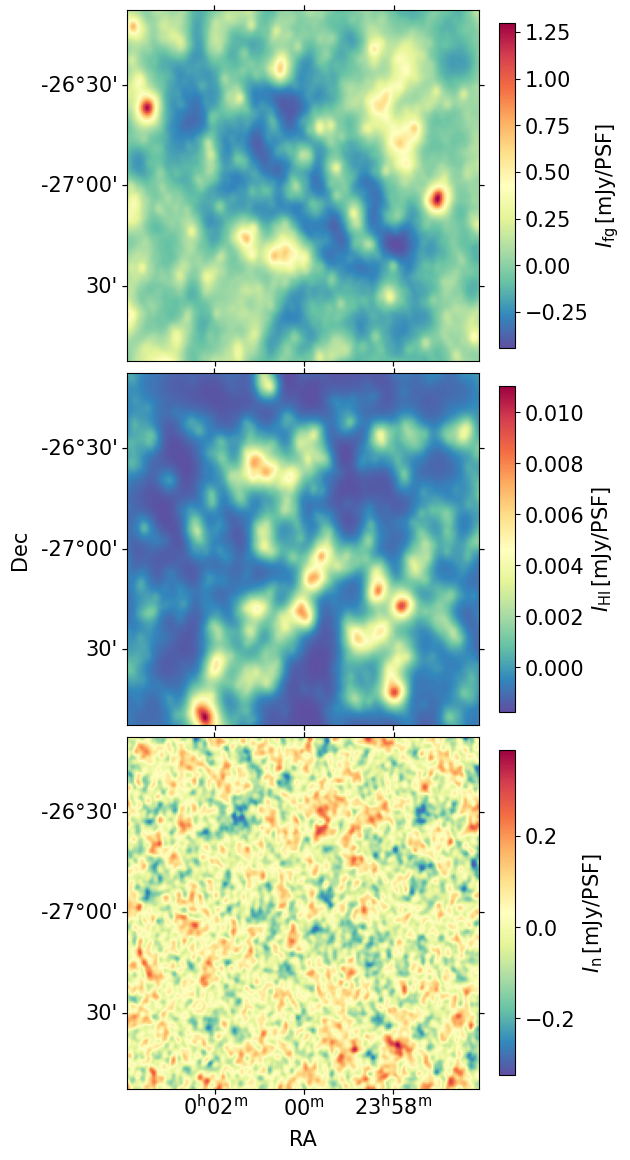}
    \caption{The output dirty images of the simulations at the central 220MHz frequency channel. The top panel shows the output dirty image of the foregrounds. The central panel shows the \hi\ image. The bottom panel shows the dirty image of the thermal noise. The images have a size of $\rm (1.5\,deg)^2$ with $200\times 200$ pixels. The images are shown in the units of mJy/PSF.}
    \label{fig:image}
\end{figure}

\subsection{Simulating systematics}
Real observations will contain a wealth of systematics, including the radio frequency interference (RFI), gain instabilities, calibration errors and more. While it is beyond the scope of this paper to properly take into account all of the systematics, we aim to simulate the effect of systematics that can lead to spectral instability in a simplistic way. The systematics are simulated using 
\begin{equation}
    V_{\rm obs}^{\rm i}(u^{\rm i},v^{\rm i},f^{\rm i}) = (1+\delta e_{f})V_{\rm true}^{\rm i}+V_{\rm TN},
\label{eq:sys}
\end{equation}
where $V_{\rm true}^{\rm i}$ is the visibility data of the $\rm i^{\rm th}$ baseline without the systematics and the thermal noise. {$V_{\rm TN}$ is the thermal noise visibility}. $\delta e_{f}$ follows a Gaussian distribution with zero mean and only depends on the frequency channel. We simulate $\delta e_{f}$  with different standard deviations from $10^{-5}$ up to $10^{-4}$. {The systematic errors are multiplied to the full visibility data before the assumed sky model subtraction.} This effect is a crude approximation for bandpass calibration error averaged across all timesteps, creating fluctuations on small frequency scales which will leak foreground power into the observation window and bias the foreground removal techniques as we discuss in Section \ref{sec:forecast}. {Note that, the calibration errors are complex and have smooth structures in frequency for \hi\ observations (see e.g. figs 2 and 3 of \citealt{2019ApJ...875...70B}). In our case, we focus on the blind removal of residual foreground after calibration and choose Gaussian errors so that the foreground scatter is present across the delay space (see Appendix \ref{sec:caveat}).}

It is worth pointing out that the 200-240 MHz frequency range hosts several prominent sources of RFI. Around 220MHz there are the RF11 and RF12 bands of digital TV (see e.g. fig. 2 of \citealt{2015PASA...32....8O}), which can be {identified through flagging algorithms (e.g. \citealt{2010MNRAS.405..155O,2019PASP..131k4507W})}. The larger end of the frequency range $\sim 240$MHz sits right next to military satellite band (242-272 MHz) which may {cause complete data loss of the entire frequency range} (see fig. 4 of \citealt{7386856}). The presence of this RFI forbids us to go below redshifts $z<5$. Overall we expect that the 200-240MHz frequency range can be observed without substantial loss of data.

\subsection{\althi\ power spectrum from the imaging route}
The image cube can be used to estimate the \hi\ power spectrum. We compute the HI power spectrum from the imaged data instead of measuring the delay power spectrum directly from the visibilities for two reasons. 
First, the cosmological quantities such as the Hubble parameter and the comoving distance have significant evolution across the large redshift bin $\Delta z= 1$, making the delay power spectrum estimation very difficult especially with regards to deconvolving w-projection kernel and primary beam attenuation. Second, if we can verify the detectability of one field in image space, we can probe larger cosmological scales through image mosaicing of overlapping fields.

\begin{figure}
    \centering
	\includegraphics[width=\columnwidth]{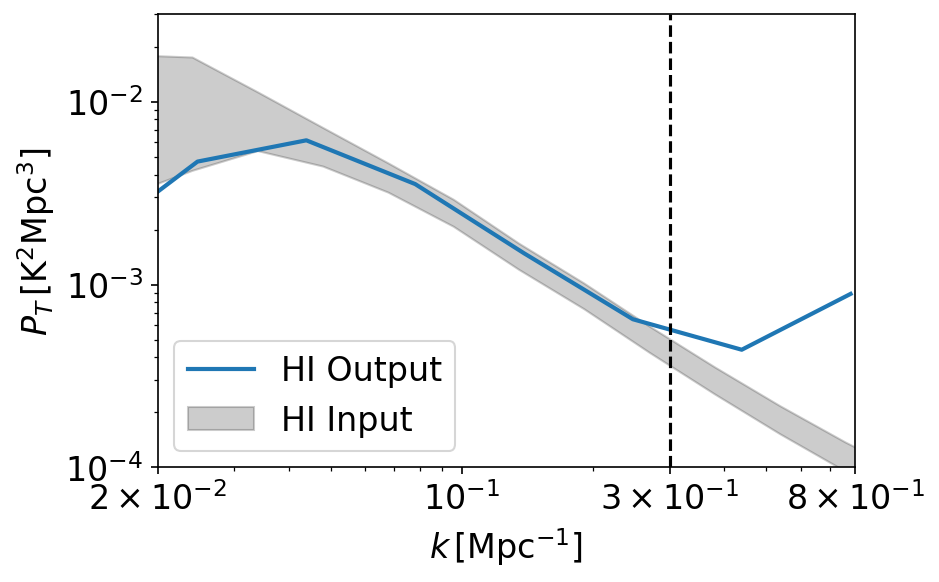}
    \caption{The \hi\ power spectrum estimated from the image cube using \hi-only visibility data (`\hi\ output'), compared against the input \hi\ power spectrum described in Section \ref{subsec:hisim} (`\hi\ input'). The vertical black dashed line corresponds to $k=0.3\,{\rm Mpc^{-1}}$ where the effects of the PSF starts to dominate.}
    \label{fig:1dps}
\end{figure}

To calculate the \hi\ power spectrum, we first transform the flux density $I(l,m,f)$ in the image cube into Fourier space brightness temperature
\begin{equation}
    \tilde{T}(\bm{k}_\perp,k_\parallel) = \int \frac{{\rm d}^3x}{{\rm \mathbf{V}}} {\rm exp}\big[-i\bm{k}\cdot \bm{x} \big]\bigg(\frac{\lambda^2}{\rm 2k_B}\bigg)^2 \frac{I(\bm{x})}{A(\bm{x})},
\label{eq:tf}
\end{equation}
where $\bm{x} = [ l\cdot {D_c}(z_{\rm c}),m\cdot D_c(z_{\rm c}), {D_c}(z_f) ]$ is the physical coordinate corresponding to the sky coordinate $(l,m)$ and observing frequency $f$. {$\mathbf{V}$ is the comoving volume of the image cube.} ${D_c}(z)$ is the comoving distance at redshift $z$. $z_{\rm c}$ is the centre of the redshift bin and $z_f = f_{21}/f-1$ is the redshift corresponding to the frequency $f$ where $f_{\rm 21}$ is the rest frequency of the 21-cm line. {$A(\bm{x})$ is the primary beam attenuation}. $\lambda$ is the observing wavelength. The transverse coordinates for each voxel are assigned assuming an effective comoving distance, which is important to ensure that the operators for residual foreground removal and Fourier transformation are commutable as we discuss in Appendix \ref{sec:est}. $\tilde{T}(\bm{k}_\perp,k_\parallel)$ is in the units of {kelvin/PSF}. The \hi\ power spectrum in 3D $k$-space is
\begin{equation}
    P_{\rm \hi}(\bm{k}_\perp,k_\parallel) = \frac{|\tilde{T}(\bm{k}_\perp,k_\parallel)|^2 }{|\widetilde{\rm PSF}(\bm{k}_\perp,f_{\rm c})|^2},
\label{eq:pscal}
\end{equation}
where $\widetilde{\rm PSF}(\bm{k}_\perp,f_{\rm c})$ is the 2-D Fourier transform of the PSF at the central frequency $f_{\rm c}$
\begin{equation}
    \widetilde{\rm PSF}(\bm{k}_\perp,f_{\rm c}) = \int {\rm d}l{\rm d}m\, {\rm exp}\big[ -2\pi i (lu+mv)\big]{\rm PSF}(l,m,f_{\rm c}).
\end{equation}

In the calculations above, several approximations have been made. The frequency evolution of the PSF is assumed to be negligible over the frequency bandwidth of the simulated observation. The physical coordinates of the voxels are assigned assuming an effective comoving distance. The flat-sky approximation is also used. While these assumptions may not be accurate enough for precision cosmology, as we show in Fig. \ref{fig:1dps}, {it can be seen that the output \hi\ power spectrum is within the $1$-$\sigma$ region of the input.} It is sufficiently accurate for studying the detectability of the signal. The scales probed are from $k\sim 0.03\,{\rm Mpc^{-1}}$, limited by the size of the image, to $k\sim 0.3\,{\rm Mpc^{-1}}$, limited by the image resolution due to the PSF. In the power spectrum results shown hereafter, a Blackman-Harris frequency taper is also applied to minimise potential leakage of foregrounds and systematics, with the details discussed in Appendix \ref{sec:est}.

\section{Quantifying the Foreground Wedge}
\label{sec:fgwedge}
In this Section, we use the image cube from \hi\ and foreground visibility data without the thermal noise to explore the limits of reducing foreground contamination. Without the thermal noise and any systematics, the \hi\ and foreground-only case showcases the best possible scenario for residual foreground removal. It helps us understand the requirements for sky modelling to enable detection and locate the observation window in the $k_\perp-k_\parallel$ plane. We particularly focus on scales of {cosmological interest} $k<0.2\,{\rm Mpc^{-1}}$, especially the largest scale that can be probed using our image cube $k\sim 0.03\,{\rm Mpc^{-1}}$. If these scales can be probed with little foreground contamination, future surveys using wide-field imaging and mosaicing can further extend the scales larger than the first Baryon Acoustic Oscillation (BAO, \citealt{1998ApJ...496..605E}) peak at $k\sim 0.04 \,{\rm Mpc^{-1}}$ to the linear scales for cosmological analysis. 

\subsection{Observation window using only foreground avoidance}
\label{subsec:avoid}

We first use the \hi-only image cube and foreground-only image cube to estimate the power spectra for the \hi\ and the foregrounds to compare them in cylindrical $k$-space. The cylindrical power spectra for the \hi\ and the foregrounds are shown in Fig. \ref{fig:hifgcy}. Comparing the ratio between the \hi\ power spectrum and the foreground power spectrum as shown in the top-left panel of Fig. \ref{fig:window}, the foreground power spectrum is larger than the \hi\ power spectrum at $k_\parallel \lesssim 0.12\,{\rm Mpc^{-1}}$, leaving no observation window at linear and BAO scales. In Fig. \ref{fig:window}, the region where foreground power dominates does not have a clear wedge structure. This is due to the fact that the bright sources in the primary beam side-lobes, which contributes mostly to the wedge structure at high $k_\perp$, are assumed to be already removed in our simulation. Without the strong foreground emissions coming from large angular extent (high delay time), the wedge structure at high $k_\perp$ no longer exists. The lack of wedge feature can also be seen from observations (e.g. LOFAR observaions shown in \citealt{2020MNRAS.493.1662M,2021MNRAS.500.2264H}). As we demonstrate in Section \ref{subsec:fgremov}, the wedge structure reappears after foreground cleaning is applied to the data. This is due to the fact that removing residual foregrounds reduces the foreground power near the pointing centre, making the foreground emission at larger angular distance comparatively brighter.

\begin{figure}
    \centering
	\includegraphics[width=\columnwidth]{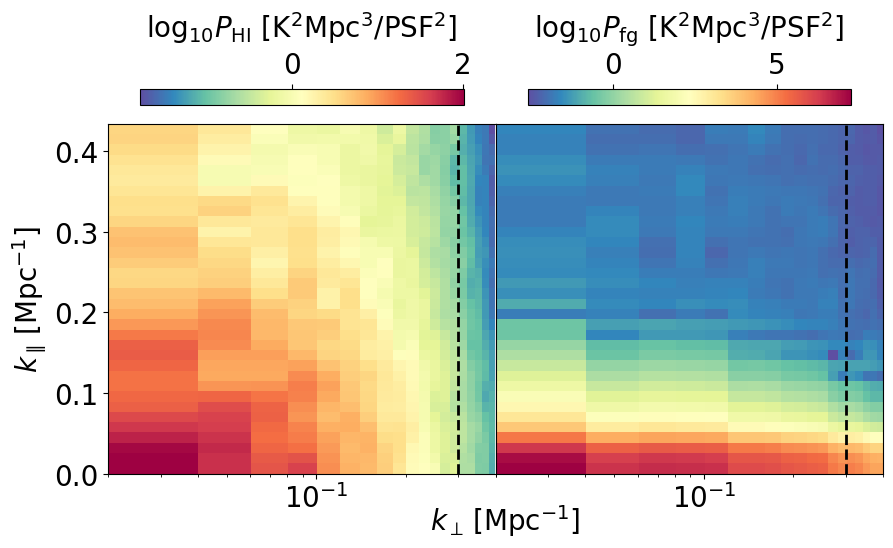}
    \caption{The cylindrical power spectra for the \hi\ (left) and the foregrounds (right) estimated from the output image cubes. Note that the PSF is not deconvolved from the power spectra and the power spectra are in the units of $\rm K^2Mpc^3/PSF^2$.}
    \label{fig:hifgcy}
\end{figure}

If we relax the $10\%$ modelling residual as described in Section \ref{subsec:exgal} to an extreme $0.1\%$, $k_\parallel \lesssim 0.05\,{\rm Mpc^{-1}}$ scales are still lost as shown in the top right panel of Fig. \ref{fig:window}, which invalidates the usage of the observations for cosmology. The result suggests that even with extreme level of calibration and sky modelling accuracy, it is unlikely that foreground avoidance can be used to measure the \hi\ power spectrum at cosmological scales at $5<z<6$ due to the weakness of the \hi\ signal at these redshifts. We can use foreground removal methods to mitigate the contamination at large scales as we show in the following sections.

\begin{figure}
    \centering
	\includegraphics[width=\columnwidth]{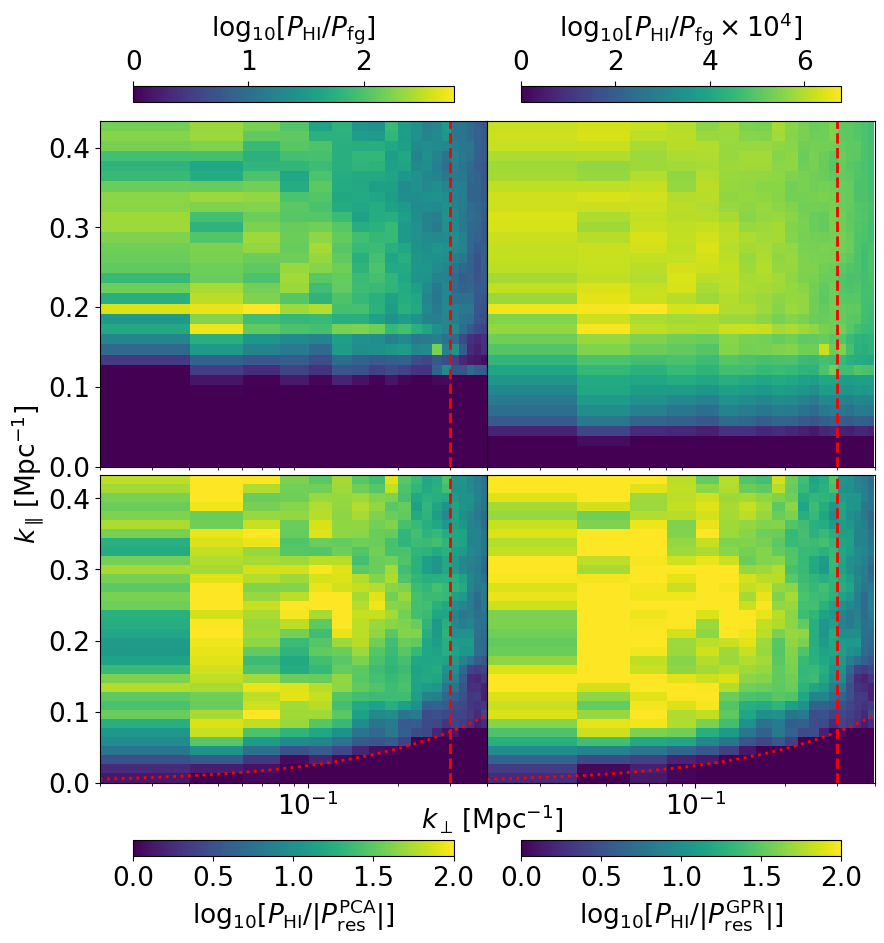}
    \caption{Top left panel: The ratio between the \hi\ power spectrum and the foreground power spectrum in cylindrical $k_\perp - k_\parallel$ space. The vertical red dashed line denotes the $k=0.3\,{\rm Mpc^{-1}}$ line where the effects of the PSF start to dominate. Top right panel: The ratio between the \hi\ power spectrum and the foreground power spectrum, with the foreground power suppressed by a factor of $10^4$. Bottom left panel: The ratio between the \hi\ power spectrum and the residual power spectrum after PCA cleaning. The red dotted line denotes the `horizon limit' $k_\parallel = 0.24 k_\perp$ calculated according to Eq. (\ref{eq:horizon}). Bottom right panel: The ratio between the \hi\ power spectrum and the residual power spectrum after GPR cleaning. In all the panels shown, values below $1$ are set to $1$ for better presentation. The darkest end of the color scale corresponding to the region where the foreground power is larger than the \hi.}
    \label{fig:window}
\end{figure}

\subsection{Residual foreground removal}
\label{subsec:fgremov}
In order to suppress foreground contamination down to the wedge and create an observation window at large scales, we explore methods of blind source subtraction to remove the residual foregrounds. We focus on two methods commonly used, namely the Principle Component Analysis (PCA, e.g. \citealt{2022MNRAS.509.2048S}) and Gaussian Process Regression (GPR, e.g. \citealt{2018MNRAS.478.3640M,2022MNRAS.510.5872S}). Following \cite{2023MNRAS.518.2971C}, with the observation window enlarged due to the foreground cleaning we can choose a criteria for the power spectrum estimation in 1D $k$-space
\begin{equation}
    k_\parallel > c_k k_\perp,
\end{equation}
where $c_k$ is a constant to be set. The value for $c_k$ can be found by iteratively testing with larger values to the point where the 1D power spectrum results converge.

We write out the general formalism for frequency-frequency covariance based foreground removal methods
\begin{equation}
    \hat{\mathbf{X}}_{\rm fg} = \hat{\mathbf{C}}_{\rm fg}\hat{\mathbf{C}}^{-1} {\mathbf{X}},
\label{eq:fgremov}
\end{equation}
where $\mathbf{X}$ is the mean-centred image cube which has dimensions of $({\rm N}_f,{\rm N}_{\rm p})$ with ${\rm N}_f$ the number of frequency channels and ${\rm N}_{\rm p}$ the number of pixels in one frequency channel. $\hat{\mathbf{C}}_{\rm fg}$ is an estimation of the covariance matrix for the foregrounds and $\hat{\mathbf{C}}^{-1}$ is the inverse of the estimation of the total data covariance. For different methods such as PCA and GPR, different choices of $\hat{\mathbf{C}}_{\rm fg}$ and $\hat{\mathbf{C}}$ are used, producing different reconstructed foregrounds which we discuss in detail in Section \ref{subsec:tnfg}.

\subsubsection{Foreground removal using PCA}
The PCA method separates the foregrounds by using the eigenvalue decomposition of the frequency-frequency data covariance matrix (e.g. \citealt{2021MNRAS.504..208C})
\begin{equation}
    \hat{\mathbf{C}}_{\rm d} = \mathbf{X} \mathbf{X}^{\rm T}/({\rm N}_{\rm p}-1),
\label{eq:dcov}
\end{equation}
 The eigenvalues and eigenvectors of the covariance matrix are then calculated. An estimation of the foregrounds can be extracted from the data matrix using 
\begin{equation}
    \hat{\mathbf{X}}_{\rm fg}^{\rm PCA} = \mathbf{A}\mathbf{A}^{\rm T}\mathbf{X},\ \mathbf{A} = [\mathbf{v}_1,...,\mathbf{v}_{\rm N_{\rm fg}}].
\label{eq:pca}
\end{equation}
Here, $\mathbf{v}_{\rm i}$ is the eigenvector corresponding to the $\rm i^{th}$ largest eigenvalue and a total of $N_{\rm fg}$ modes are removed. To link it to Eq. (\ref{eq:fgremov}), we can rewrite Eq. (\ref{eq:pca}) as
\begin{equation}
    \hat{\mathbf{X}}_{\rm fg}^{\rm PCA} = \big(\mathbf{A}\mathbf{A}^{\rm T} \hat{\mathbf{C}}_{\rm d}  \big) \big(\hat{\mathbf{C}}_{\rm d}\big)^{-1} \mathbf{X},
\label{eq:pcaform}
\end{equation}
where it is straightforward to see that, in the case of PCA, $\hat{\mathbf{C}}_{\rm PCA}=\hat{\mathbf{C}}_{\rm d}$ and $\hat{\mathbf{C}}_{\rm fg}^{\rm PCA} =\mathbf{A}\mathbf{A}^{\rm T} \hat{\mathbf{C}}_{\rm d}$.

In our case, the eigenvalues of the data covariance reach a plateau after the third eigenvalue, suggesting that $\rm N_{\rm fg}=3$ is a good choice for cleaning the foregrounds and avoiding overcleaning the signal. The ratio between the \hi\ power spectrum and the residual power spectrum after cleaning is shown in the bottom left panel of Fig. \ref{fig:window}. Throughout the paper, the residual power spectrum is defined as the power spectrum of the residual foreground image $\mathbf{X}_{\rm res} = \mathbf{X}_{\rm fg} - \hat{\mathbf{X}}_{\rm fg}$, where $\mathbf{X}_{\rm fg}$ is the image of the input foregrounds and $\hat{\mathbf{X}}_{\rm fg}$ is the removed foreground by either PCA or GPR. 

Comparing Figs \ref{fig:hifgcy} and Fig. \ref{fig:window}, the cleaning efficiently enlarges the observation window at small $k_\perp$. If the foreground contamination is optimally mitigated, the foreground wedge can be located using the `horizon limit' \citep{2014PhRvD..90b3018L}
\begin{equation}
    c_{k}^{\rm h} = \frac{H(z)D_{c}(z)\theta_0}{c(1+z)},
\label{eq:horizon}
\end{equation}
where {$H(z)$ is the Hubble parameter,} $c$ is the speed of light and $\theta_0$ is the angular extent of the instrument beam. As a crude approximation we choose $\theta_0 = 2\sqrt{\Omega_{\rm beam}}/\pi$ where $\Omega_{\rm beam}$ is the integrated primary beam which gives $c_k^{\rm h} = 0.24$. From the bottom-left panel of Fig. \ref{fig:window}, we can see that the foreground wedge is close to the horizon limit which is marked by the red dotted line, showing that the foreground cleaning is efficient. Iteratively increasing the threshold we find that the 1D power spectrum converges at $c_k = 0.3$, which we use from now on in this paper.

\subsubsection{Foreground removal using GPR}
GPR constructs the foreground component by fitting parameterized kernels to the data covariance. Suppose we have the \hi\ kernel $\mathbf{K}_{\rm \hi}$, the foreground kernel $\mathbf{K}_{\rm fg}$ and the thermal noise kernels $\mathbf{K}_{\rm n}$ fitted, then the estimated foreground can be written as (e.g. \citealt{2018MNRAS.478.3640M})
\begin{equation}
    \hat{\mathbf{X}}_{\rm fg}^{\rm GPR} = {\mathbf{K}}_{\rm fg}\big(\mathbf{K}_{\rm fg}+\mathbf{K}_{\rm n}+\mathbf{K}_{\rm \hi}\big)^{-1} {\mathbf{X}}.
\label{eq:gpr}
\end{equation}
It is straightforward to see that, in the case of GPR, $\hat{\mathbf{C}}_{\rm GPR}=\mathbf{K}_{\rm fg}+\mathbf{K}_{\rm n}+\mathbf{K}_{\rm \hi}$ and $\hat{\mathbf{C}}_{\rm fg}^{\rm GPR} =\mathbf{K}_{\rm fg}$. 

\begin{figure}
    \centering
	\includegraphics[width=\columnwidth]{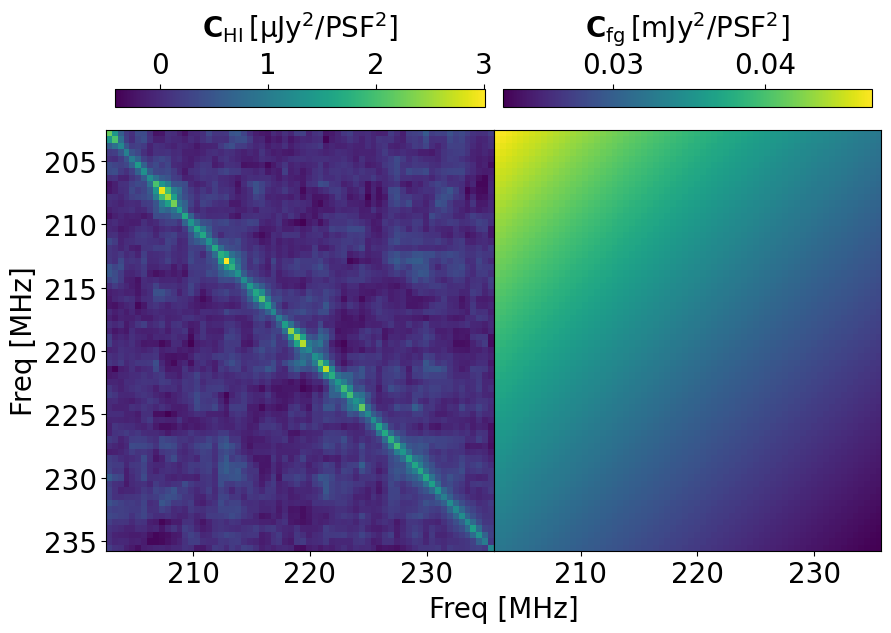}
    \caption{The frequency-frequency covariance matrices for the \hi\ ($\mathbf{C}_{\rm \hi}$) on the left and the foregrounds ($\mathbf{C}_{\rm fg}$) on the right. The covariance matrices are calculated from \hi-only and foreground-only image cubes following Eq. (\ref{eq:dcov}).}
    \label{fig:hifgcov}
\end{figure}

The \hi\ and the foreground covariance matrices are shown in Fig. \ref{fig:hifgcov} for reference. The \hi\ covariance is highly diagonal, due to the discrete and uncorrelated nature of the \hi\ along the line-of-sight. On the other hand, the foreground covariance is smooth and shows a clear spectral feature along the frequency direction, corresponding to the negative spectral indices of the radio sources. Due to the spectral evolution of the foreground covariance, the conventional choice of a Mat{\'e}rn kernel \citep{matérn1966spatial} does not describe the foreground covariance well. Instead, we use Markov chain Monte Carlo (MCMC) to fit the kernels using the following steps:
\begin{itemize}
    \item 
    In each step, a random value $\sigma_{\rm n}$ is sampled and a diagonal kernel $ \mathbf{K}_{\rm n} = \sigma_{\rm n}^2 \delta_{\rm ij}^{\rm K}$ is calculated where $\delta^{\rm K}$ is the Kronecker delta. In this section, $\mathbf{K}_{\rm n}$ is the \hi\ kernel. Following \cite{2022MNRAS.510.5872S}, in Section \ref{sec:forecast} when thermal noise is included, $\mathbf{K}_{\rm n}$ is the sum of the \hi\ and the thermal noise covariance matrices.
    \item
    The total data covariance matrix is then subtracted by the diagonal kernel $ \mathbf{K}_{\rm n}$. A third-order polynomial fitting is then performed on every row of the subtracted result, creating a fitted kernel $\mathbf{K}_{\rm fit}$. The kernel is then symmetrised to get the foreground kernel $\mathbf{K}_{\rm fg} = (\mathbf{K}_{\rm fit}+\mathbf{K}_{\rm fit}^{\rm T})/2$.
    \item
    {The parameters for the kernels are then fitted by maximizing the log-marginal likelihood ${\rm log}p = -(\mathbf{X}^{\rm T}\mathbf{K}^{-1}\mathbf{X}+{\rm log}|\mathbf{K}|+n{\rm log}2\pi)/2$, where $n$ is the number of data points sampled and $\mathbf{K}$ is the sum of the kernels $\mathbf{K} = \mathbf{K}_{\rm fg}+\mathbf{K}_{\rm n}$.}
    \item
    The MCMC fitting is then performed with 20 random walkers with 2000 iterations to make sure the chains converge. The initial guess of $\sigma_{\rm n}$ is taken to be the square root of the trace of the data covariance. The final kernels are the 50\% percentile of the $ \mathbf{K}_{\rm n}$ and the $ \mathbf{K}_{\rm fg}$ samples in the chains excluding the first 100 steps.
\end{itemize}

\begin{figure}
    \centering
	\includegraphics[width=\columnwidth]{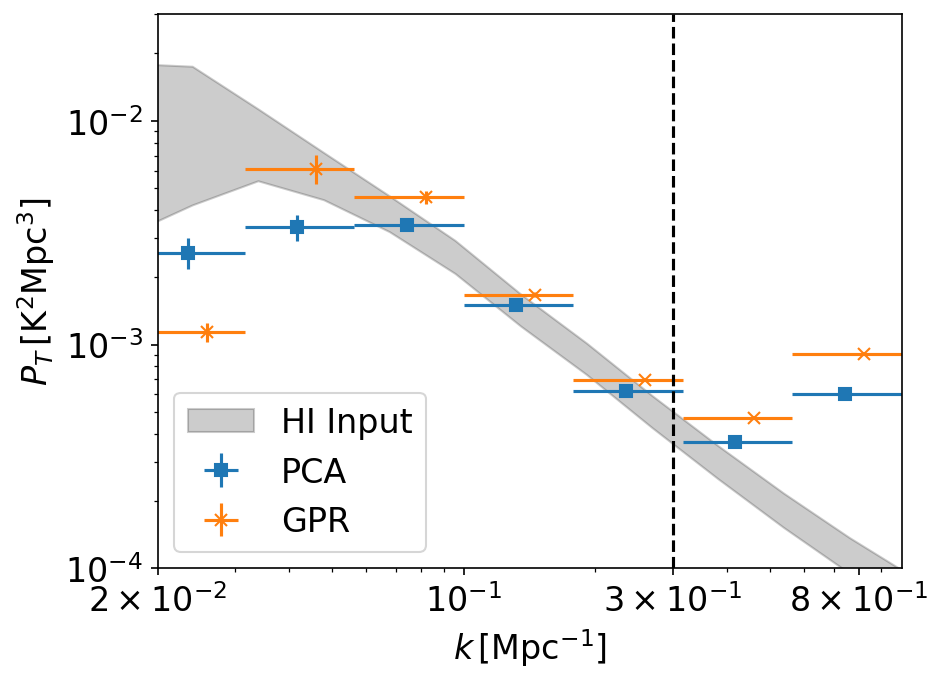}
    \caption{The 1D \hi\ power spectrum results from image cubes of \hi\ and foreground using foreground removal methods measured in the $k_\parallel>0.3k_\perp$ regions. The blue data points shows\ the results from PCA and the yellow points show the results from GPR. The vertical dashed line denotes $k=0.3\,{\rm Mpc^{-1}}$ where the effects of PSF start to dominate. The centres of the $k$-bins are misplaced by 5\% for presentation.}
    \label{fig:1dnonoise}
\end{figure}

Note that after GPR cleaning, a bias correction can be applied as shown in \cite{2018MNRAS.478.3640M}. We follow the quadratic estimator formalism of \cite{2021MNRAS.501.1463K} and show in Appendix \ref{sec:est} that the bias correction term in our case is negligible. The resulting foreground residual power spectrum compared to the \hi\ power spectrum is shown in the bottom right panel of Fig. \ref{fig:window}. Comparing the foreground wedge in the GPR case with the horizon limit and with the PCA case, we can see that in the absence of thermal noise, GPR is slightly more efficient in cleaning the foregrounds and both methods do well enough to enable the detection of the \hi\ at large scales $k<0.1\,{\rm Mpc^{-1}}$. At the largest spatial scales of the image, there is negative residual power from overcleaning. The differences between these two methods are discussed later in Section \ref{subsec:tnfg}.

The success of the methods in cleaning the foregrounds indicates that we can measure the \hi\ power spectrum from the SKA-Low observation at $5<z<6$, as we show using the 1D power spectrum in Fig. \ref{fig:1dnonoise}. As mentioned, both methods can enable the measurements of the \hi\ power spectrum from $k\sim 0.05 \,{\rm Mpc^{-1}}$ up to $k\sim 0.3\,{\rm Mpc^{-1}}$.

\section{Forecasts for SKA-Low}
\label{sec:forecast}
In this section, we further explore the detectability of the \hi\ power spectrum for SKA-Low observations by including different levels of thermal noise in the simulation. In particular, to enable the measurement of the \hi\ power spectrum, the robustness of the foreground removal methods in the presence of the thermal noise must be tested. Furthermore, we simulate systematics by generating stochastic errors along the frequency direction to test the limits of level of systematics allowed.

\subsection{Robust foreground cleaning with low SNR}
\label{subsec:tnfg}
In Section \ref{subsec:fgremov}, we show that the foreground removal methods can suppress the foreground wedge to the horizon limit. However, this result is based on the fact that the empirical data covariance is `clean', i.e. the covariance is purely a combination of the \hi\ and the foregrounds. Therefore, the distinctive features of the \hi\ can be extracted from the signal using PCA and GPR. In reality, the data covariance is likely to contain a high level of thermal noise as well as systematics, making it difficult to construct the covariance of the foregrounds. We test PCA and GPR in the presence of different levels of thermal noise. As described in Eq. (\ref{eq:sigman}) in Section \ref{subsec:skytoim}, we simulate the thermal noise for the 12h tracking and rescale it to match 360, 480 and 600 h of integration time.

\begin{figure}
    \centering
	\includegraphics[width=\columnwidth]{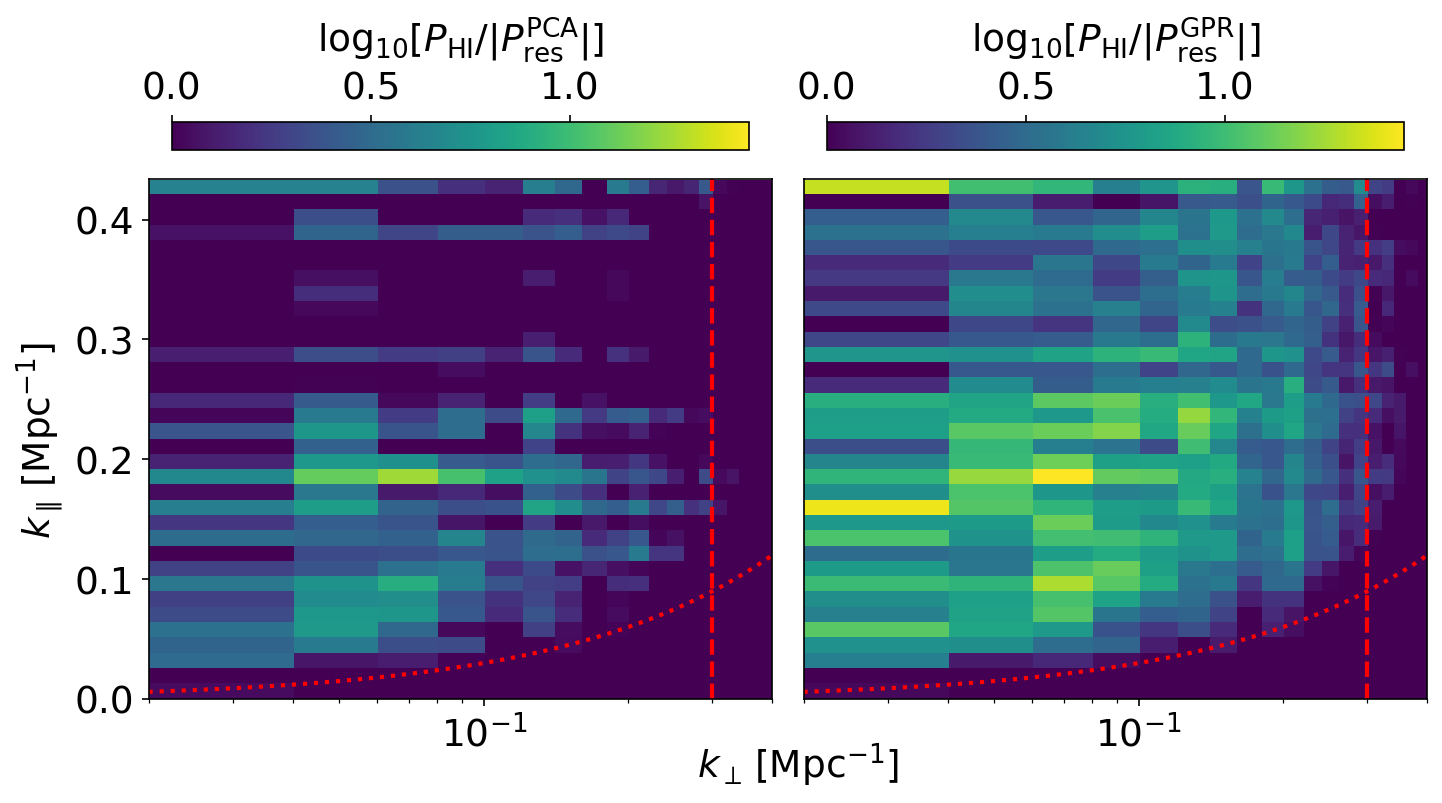}
    \caption{Left panel: The ratio between the \hi\ power spectrum and the residual foreground power spectrum in cylindrical $k_\perp - k_\parallel$ space using the PCA cleaning. Right panel: The same with the left panel except the residual is obtained using the GPR cleaning. All panels shown have values below $1$ set to $1$ to separate the observation window from the foreground wedge. The vertical red dashed line denotes the $k=0.3\,{\rm Mpc^{-1}}$ line where the effects of the PSF start to dominate. The red dotted line denotes the boundary for the observation window $k_\parallel = 0.3 k_\perp$. }
    \label{fig:w30n}
\end{figure}

We first show the results for the 360 h case and compare the effects of foreground removal methods. The PCA and GPR routines are kept the same as in Section \ref{subsec:fgremov} with the observation window $c_k = 0.3$. The ratio between the underlying \hi\ power spectrum and the foreground residual after removal in cylindrical $k$-space is shown in Fig. \ref{fig:w30n}. In contrast with the results shown in Fig. \ref{fig:window}, the amplitude of the residual power increases significantly. For the PCA case, the observation window is heavily contaminated by the foregrounds while the contamination is less severe in GPR. Note that, this difference is not visible in the residual image cube as we show in Fig \ref{fig:resim}. The amplitude of the fluctuation of the residual is roughly the same with no indications of the different levels of foreground contamination.

\begin{figure}
    \centering
	\includegraphics[width=\columnwidth]{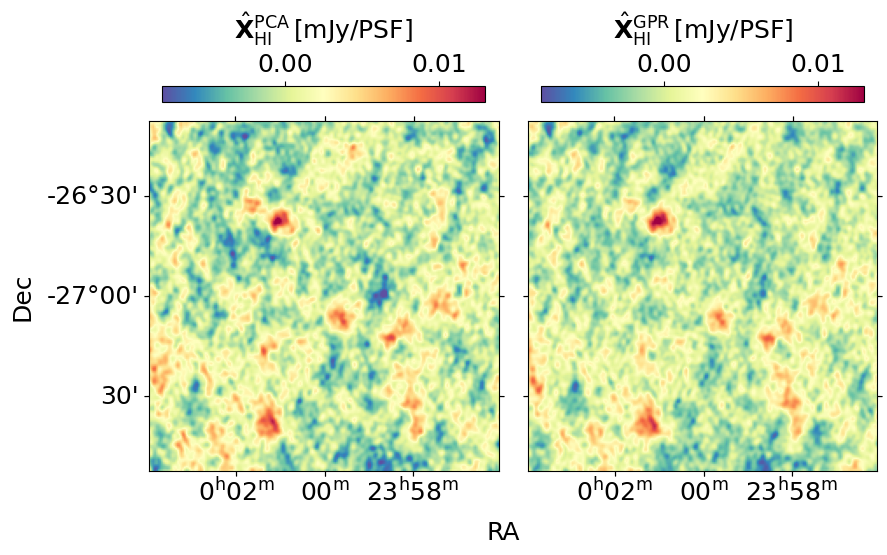}
    \caption{Left panel: The estimated \hi\ image $\hat{\mathbf{X}}_{\rm \hi}$ defined in Eq. (\ref{eq:esthi}) using the PCA cleaning at the central 220MHz frequency channel. Right panel: The same with the left panel except the residual is obtained using the GPR cleaning. The color scales of the images are set to range from -0.11 to 0.13 mJy/PSF for fair comparisons. The residual images obtained from PCA and GPR are similar with each other, yet the level of foreground leakage differs significantly as shown in Fig. \ref{fig:w30n}.}
    \label{fig:resim}
\end{figure}

The difference between PCA and GPR can be seen using the formalism in Section \ref{subsec:fgremov}. Comparing Eq. (\ref{eq:pcaform}) and Eq. (\ref{eq:gpr}), we can see that GPR uses the fitting result to obtain smooth kernels of the \hi\ and the foregrounds for cleaning. On the other hand, PCA directly operates on the total data covariance, which contains a fluctuation around zero in the non-diagonal elements because of the thermal noise. The fluctuation of the thermal noise leads to small scale oscillations in the residual covariance. For comparison, we calculate the covariance of the `estimated' \hi, i.e. the total image subtracted by the removed foreground and the noise component 
\begin{equation}
    \hat{\mathbf{X}}_{\rm \hi } ={\mathbf{X}}_{\rm d} - {\mathbf{X}}_{\rm n} - \hat{\mathbf{X}}_{\rm fg }.
\label{eq:esthi}
\end{equation}

Comparing the covariance of $\hat{\mathbf{X}}_{\rm \hi}^{\rm PCA}$ and $\hat{\mathbf{X}}_{\rm \hi}^{\rm GPR}$ as shown in Fig. \ref{fig:cov30n}, we can see that while the amplitude of the covariance is roughly the same and close to the true \hi\ shown in the left panel of Fig. \ref{fig:hifgcov}, the PCA case has large fluctuations across the frequency channel, leading to the stripe-like features in the frequency-frequency covariance matrix. While this fluctuation is also present in GPR, its amplitude is much smaller and the dominating component is still the diagonal \hi\ covariance. For PCA, however, this fluctuation introduces a small-scale fluctuation that spills foreground power into the observation window, resulting in severe signal loss at all scales including scales where the foreground power is originally already lower than the \hi\ as shown in the upper left panel of Fig. \ref{fig:window}. To further illustrate the small-scale contamination, we calculate the covariance matrices for the residual foreground $\hat{\mathbf{X}}_{\rm res }$ for PCA and GPR and compare them with the \hi\ covariance as shown in Fig. \ref{fig:covdiff}. Both methods have clear foreground residual structure over large frequency scales. However, the PCA residual has a much larger small-scale fluctuation with the amplitude larger than the diagonal \hi. The small-scale fluctuation results in severe contamination in high $k_\parallel$ modes inside the observation window as shown in Fig. \ref{fig:w30n}.

\begin{figure}
    \centering
	\includegraphics[width=\columnwidth]{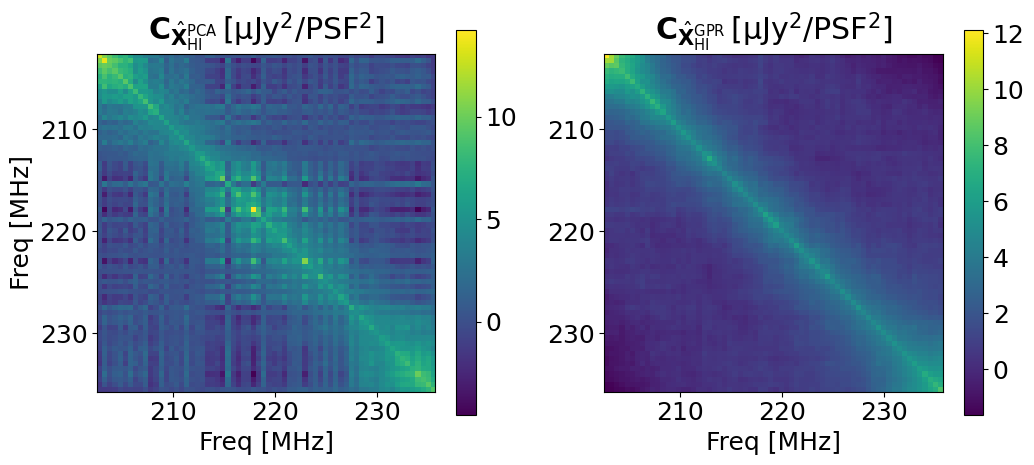}
    \caption{Left panel: The frequency-frequency covariance of the `estimated' \hi\ image $\hat{\mathbf{X}}_{\rm \hi }$ obtained using the PCA cleaning. Right panel: The same with the left panel except the residual is obtained using the GPR cleaning.}
    \label{fig:cov30n}
\end{figure}
\begin{figure}
    \centering
	\includegraphics[width=\columnwidth]{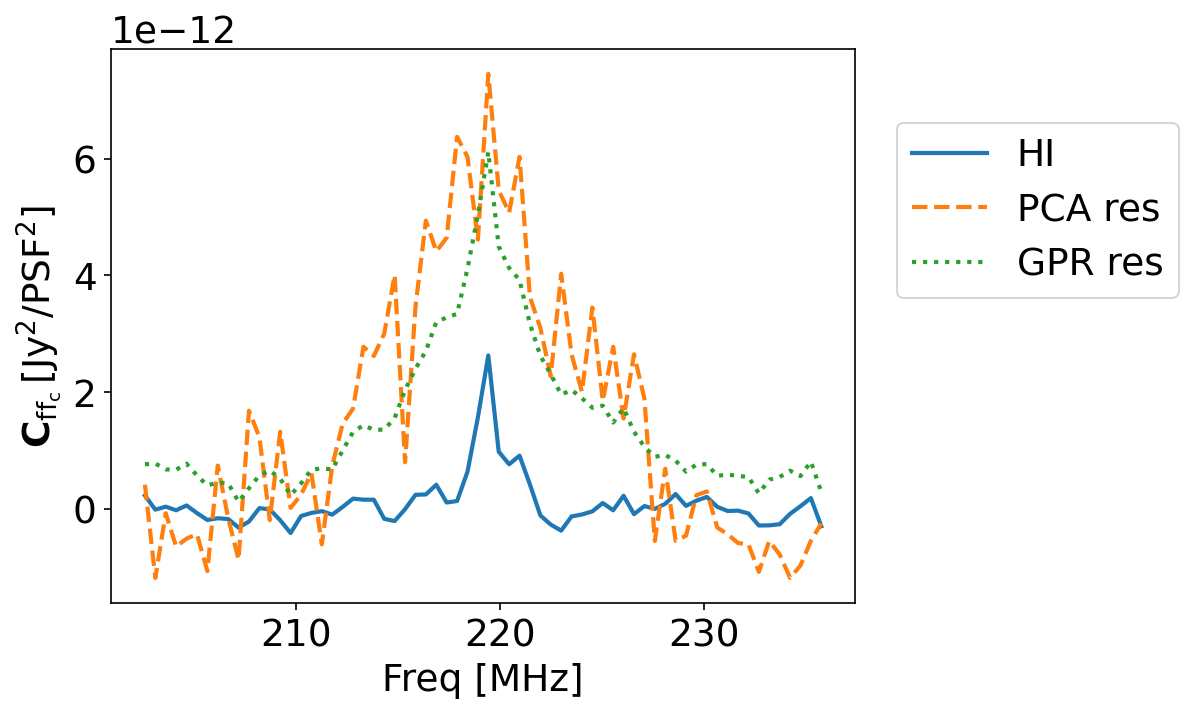}
    \caption{The frequency-frequency covariance of the residual foreground image $\hat{\mathbf{X}}_{\rm res}$ obtained using the PCA cleaning (`PCA res') and the GPR cleaning (`GPR res') for the central row $\mathbf{C}_{\rm ff_c}$ with $\rm f_c = 220\,$ MHz. The \hi\ covariance is shown in blue solid line (`HI') for reference.}
    \label{fig:covdiff}
\end{figure}
\begin{figure}
    \centering
	\includegraphics[width=\columnwidth]{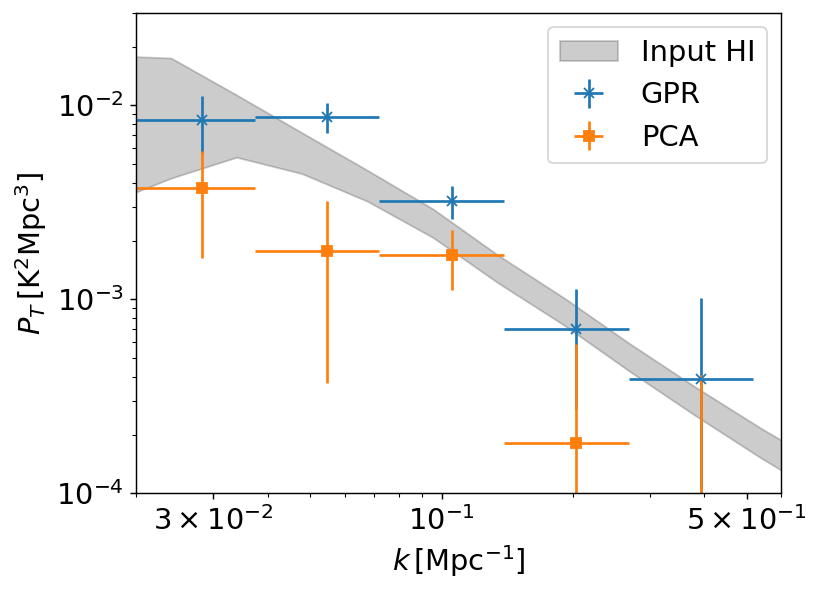}
    \caption{The 1D \hi\ power spectrum measurements with 360 h of integration time after residual foreground cleaning. The error bars on the horizontal axis denote the width of the $k$-bins and the error bars on the vertical axis denote the errors of the bandpower estimation. The results for GPR are shown in the shape of cross (`GPR') and the results for PCA are shown in the shape of squares (`PCA'). The shaded region denotes the input \hi\ power spectrum (`Input HI'). }
    \label{fig:hi30n}
\end{figure}

When comparing PCA and GPR, we assume full knowledge of the true \hi, thermal noise and foregrounds in our simulation to perform quality checks on the foreground removal methods. It is important to note that the foreground removal and power spectrum estimation routines do not rely on knowing the underlying components. The foreground removal is performed blindly and the \hi\ power spectrum is estimated by subtracting a thermal noise covariance as discussed in Appendix \ref{sec:est}. We choose logrithmically distributed k-bins from $0.01$ to $1\,$ $\rm Mpc^{-1}$ with $\Delta[{\rm log}( k/{\rm Mpc^{-1}})] = 0.25$ and show the resulting 1D power spectrum for 360 h of integration time for both PCA and GPR in Fig. \ref{fig:hi30n}. Throughout this paper, the error bars on the 1D power spectrum are calculated by calculating the sampling variance of the 3D powers that fall into the 1D $k$-bins. The resulting measurement errors on the power spectrum are
\begin{equation}
    \Delta P (k_i) = \frac{{\rm std}[P(k\in k_i)]}{\sqrt{N_{\rm modes}^i}},
\end{equation}
where ${\rm std}[P(k\in k_i)]$ denotes the standard deviation among the 3D powers that belongs in the $i^{\rm th}$ $k$-bin and $N_{\rm modes}^i$ denotes the number of $k$-points in the $i^{\rm th}$ $k$-bin.

As shown in Fig. \ref{fig:hi30n}, the foreground contamination leads to overestimation for the GPR case from $k\sim 0.03$ to $0.3\,{\rm Mpc^{-1}}$. The severe contamination of foregrounds results in signal loss on most scales for the PCA case. In conclusion, we find that in the presence of high thermal noise, PCA induces foreground contamination into the observation window due to the small-scale fluctuation in the data covariance matrix. On the other hand, GPR does not introduce sizable foreground leakage into the observation window and mitigates the foregrounds to enable the measurements of the \hi\ power spectrum at large scales.

\begin{figure}
    \centering
	\includegraphics[width=\columnwidth]{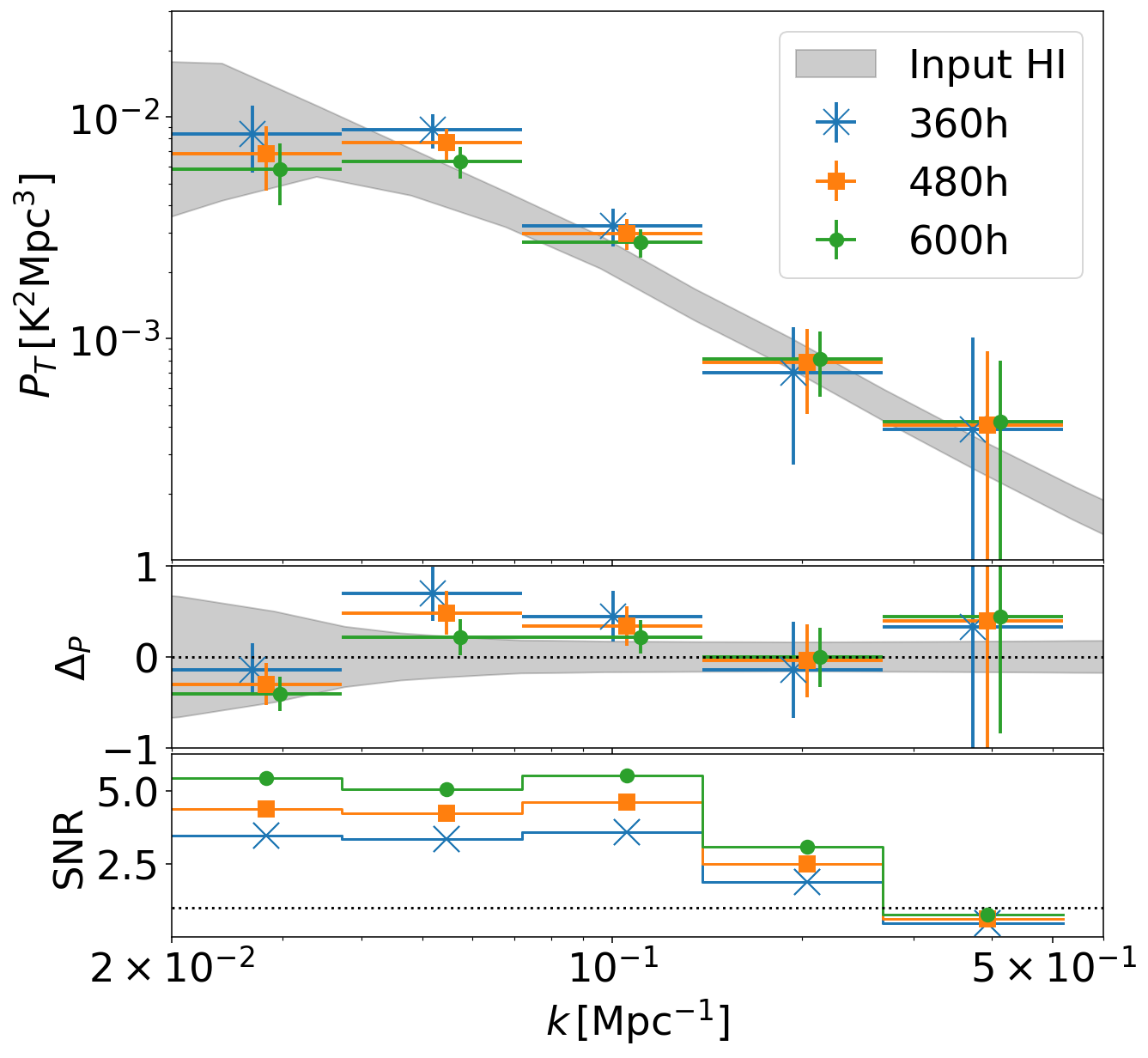}
    \caption{Top panel: The 1D \hi\ power spectrum measurements with 360, 480 and 600 h of integration time after GPR foreground cleaning. The error bars on the horizontal axis denote the width of the $k$-bins and the error bars on the vertical axis denote the errors of the bandpower estimation. The shaded region denotes the input \hi\ power spectrum (`Input HI'). The centres of the k-bins for the 360 and 600 h cases are misplaced by 5\% in $k$-direction for better presentation. Central panel: The fractional difference between the estimated \hi\ power spectrum and the underlying \hi\ input $\Delta_P = (\hat{P}_{\rm \hi}-{P}_{\rm \hi})/{P}_{\rm \hi}$. The black dotted line denotes $\Delta_P = 0$. Bottom panel: The SNR of the measurements. The black dotted line denotes SNR=1.}
    \label{fig:forecast}
\end{figure}

Using GPR, we present our forecasts for the \hi\ power spectrum measurement for SKA-Low observations of the EoR0 field assuming 360, 480 and 600 h of integration time in Fig. \ref{fig:forecast}. The power spectrum results converge to the input \hi\ as the noise level decreases. For 360 h of integration time, all bandpower measurements are within the 1-$\sigma$ error of the input \hi\ with the bandpower at $k\sim 0.05\,{\rm Mpc^{-1}}$ slightly overestimated due to foreground contamination. While not shown, we also tested that decreasing the integration time to 250 h results in the bias exceeding the 1-$\sigma$ error. We conclude that the integration time of one field needs to be greater than 250 h to enable unbiased measurements of the \hi\ power spectrum. In the case of 480 h, the \hi\ power spectrum can be measured with $\sim 3$ signal-to-noise ratio from $k\sim 0.03$ to $0.3\,{\rm Mpc^{-1}}$. Further increasing the integration time to 600 h, we find that the bias further decreases and the error bar scales as $\sqrt{t_{\rm int}}$, suggesting that the thermal noise is the dominant source of the measurement errors.

\subsection{Impact of systematics on foreground removal}
\label{subsec:system}
Interferometric observations contain various systematics, such as RFI, gain fluctuations, calibration errors, etc. These systematics impact the \hi\ power spectrum measurement in various ways. For example, the data loss coming from RFI requires inpainting or novel Fourier transform methods which leaves residuals in the power spectrum \citep{2016ApJ...818..139T,2023MNRAS.520.5552P}. Imperfect calibrations leaks the foreground power into the observation window \citep{2016MNRAS.461.3135B}. Gain and phase errors contribute to the foreground contamination in the \hi\ power spectrum \citep{2022MNRAS.515.4020M}. As a proof of concept, we are aiming to give a qualitative assessment of the impact of the systematics. {Following Eq. (\ref{eq:sys}), we set ${\rm std}(\delta e_{f})$ to $10^{-5}$, $5\times 10^{-5}$, and $10^{-4}$ to check the resulting power spectrum estimation.} All foreground removal and power spectrum estimation steps are kept the same as Section \ref{subsec:tnfg} and we choose the integration time to be 600 h to isolate the impact of the systematics. Previous literature suggest that $<10^{-5}$ level of systematic error is needed for the measurement \citep{2016MNRAS.461.3135B,2022MNRAS.515.4020M}. Note that, however, as we simulate the systematics as a random error on the true signal, it acts as a small frequency-scale fluctuation on the residual foregrounds which can be partially mitigated. Therefore, using methods such as GPR, we can still recover the observation window with the level of systematics higher than $10^{-5}$.

\begin{figure}
    \centering
	\includegraphics[width=\columnwidth]{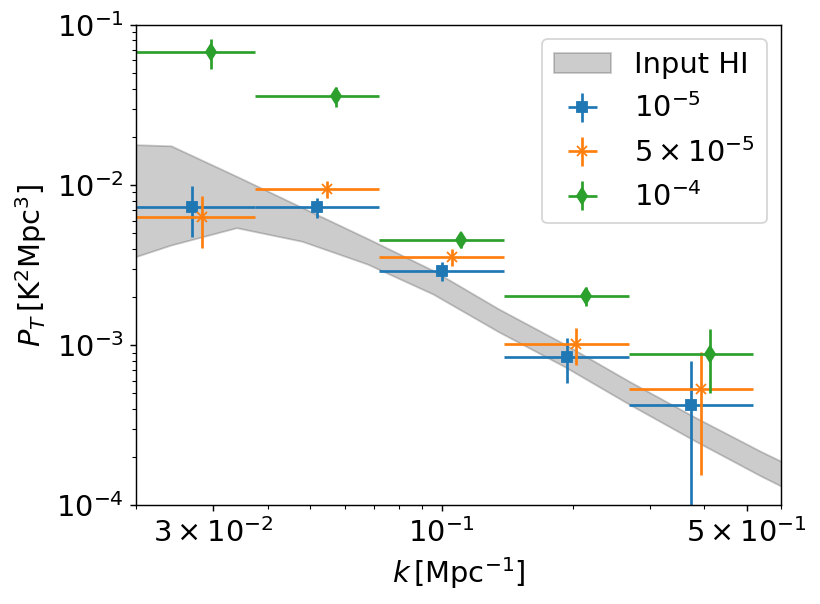}
    \caption{The \hi\ power spectrum measured from the residual foreground removed image cube as described in Section \ref{subsec:system} for different levels of systematics ($10^{-5}$, $5\times 10^{-5}$, and $10^{-4}$). The error bars on the horizontal axis denote the width of the $k$-bins and the error bars on the vertical axis denote the errors of the bandpower estimation. The shaded region denotes the input \hi\ power spectrum (`Input HI'). The centres of the k-bins for the $10^{-5}$ and $10^{-4}$ systematic effect cases are misplaced by 5\% in $k$-direction for better presentation. }
    \label{fig:hisys}
\end{figure}

In Fig. \ref{fig:hisys}, we show the \hi\ power spectrum measured from the $k_\parallel >0.3k_\perp$ window with the signal perturbed by the $10^{-5}$, $5\times 10^{-5}$, and $10^{-4}$ systematic error. For a very small level of $10^{-5}$ systematic effects, the GPR foreground removal method is unaffected by the systematics and removes the residual foreground sufficiently. As we increase the level of systematics to $5\times 10^{-5}$, the foreground starts to leak into the observation window especially at small scales, leading to overestimation of the \hi\ power spectrum. Increasing the systematics to $10^{-4}$ the contamination becomes severe and leads to biased estimation of the \hi\ power spectrum at all scales.

\begin{figure}
    \centering
	\includegraphics[width=\columnwidth]{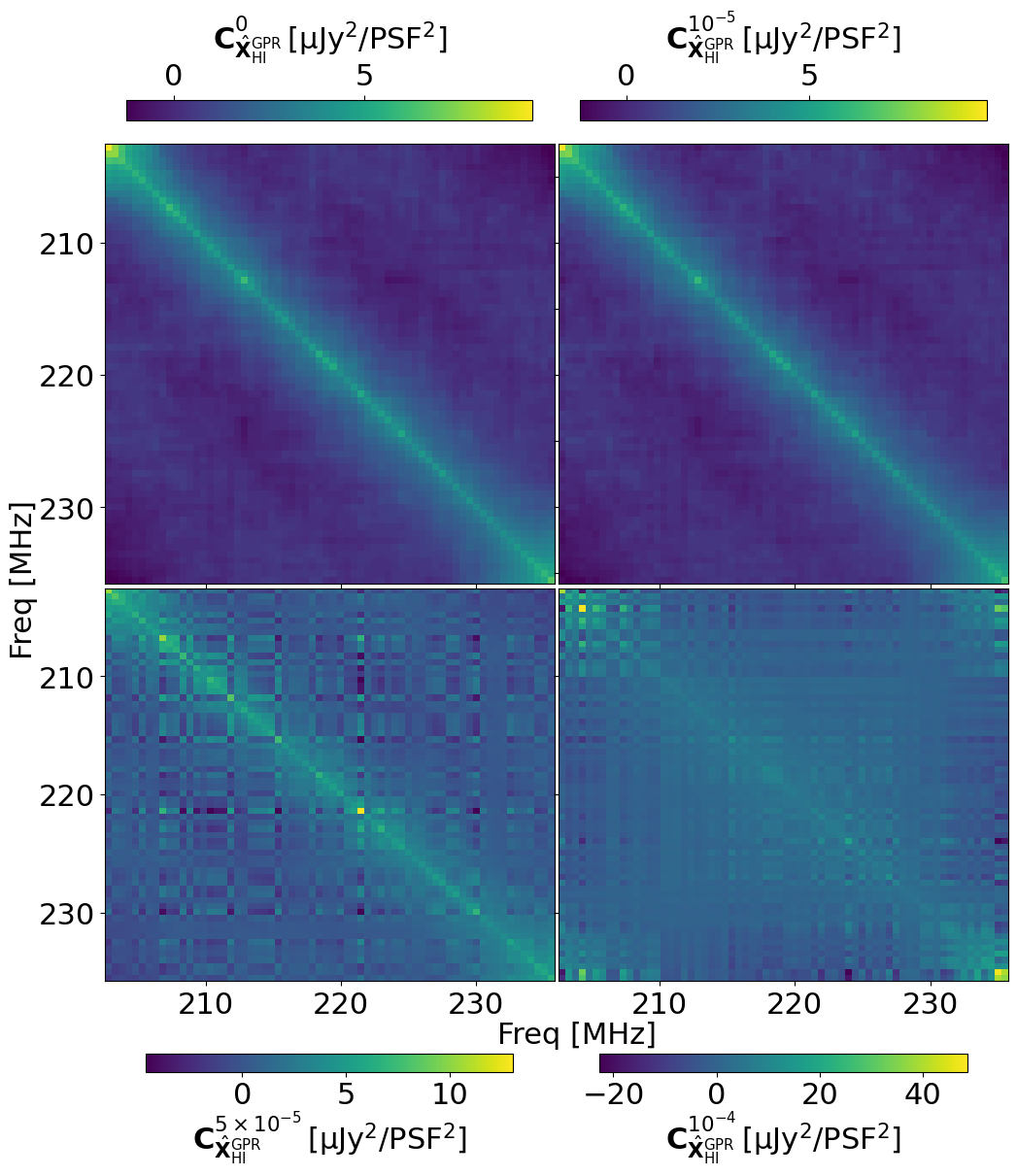}
    \caption{Top left panel: The frequency-frequency covariance matrix $\rm \mathbf{C}_{\hat{\mathbf{X}}_{\rm \hi}^{\rm GPR}}$ of the `estimated' \hi\ image $\hat{\mathbf{X}}_{\rm \hi }$ obtained using the GPR cleaning with no systematics. Top right panel: The same covariance matrix with the left panel except the simulation includes systematic effects with $10^{-5}$ fluctuations. Bottom left panel: The same covariance matrix with $5\times 10^{-5}$ systematic effects. Bottom right panel: The same covariance matrix with $10^{-4}$ systematic effects. }
    \label{fig:covsys}
\end{figure}

Similar to Section \ref{subsec:tnfg}, we can use the covariance matrices to show that the systematic effects break the frequency smoothness of the foreground, leading to biased foreground removal results. The covariance matrices of the `estimated' \hi\ in presence of different levels of systematics are shown in Fig. \ref{fig:covsys}. When no significant systematic effects are included as shown in the top panels, the reconstructed \hi\ covariance matrix is largely diagonal, suggesting that no sizeable foreground leakage is present. However, as we increase the level of systematics to $5\times 10^{-5}$, the small scale stripes similar to the ones discussed in Section \ref{subsec:tnfg} appear. For the $5\times 10^{-5}$ case, we can see that the diagonal component is still dominant and indeed as shown in Fig. \ref{fig:hisys} the \hi\ power spectrum is still accurately measured. Increasing the level of systematics to $10^{-4}$, we can see that the covariance is completely dominated by the contamination from the systematics, leaving no observation window for the \hi\ at all. In conclusion, the level of residual systematics needs to be contained at $<10^{-4}$ and ideally $\lesssim 5\times 10^{-5}$ for accurate measurement of the \hi\ power spectrum.

\section{Conclusions}
\label{sec:conclusion}
In this paper, we present the first proof of concept for measuring the \hi\ power spectrum at $5<z<6$ using SKA-Low. We have presented an end-to-end simulation and data analysis pipeline, generating the sky signal, the interferometric observation, performing the imaging and the power spectrum estimation. We use the pipeline to generate realistic simulations consistent with deep observations of the EoR0 field using SKA-Low and test foreground mitigation methods to present our forecasts for future SKA-Low observations.

We start by simulating the input sky signal including the \hi\ and the foregrounds. Galactic foregrounds are generated based on templates from observed maps of the radio sky and extrapolated to the frequency range of our interests. We use a realistic radio source catalogue to simulate the extragalactic radio sources. The \hi\ clustering signal is generated by using large-volume, realistic dark matter halo simulations with an \hi\ HOD inpainting. Generating the sky signal with different levels of foreground residuals compared with the underlying \hi\ signal, we find that:
\begin{itemize}
    \item
    {Assuming a realistic amplitude for foreground residuals after sky model subtraction to be at $\sim 80$mJy in the image cube}, the foregrounds reside mainly at low $k_\parallel \lesssim 0.1\, {\rm Mpc^{-1}}$, leaving an observation window at high $k_\parallel$ for estimation of the \hi\ power spectrum. {Residual foregrounds need to be subtracted using blind source separation methods to enable the measurement of the \hi\ power spectrum at large cosmological scales $k<0.1\,{\rm Mpc^{-1}}$.}
    \item
    Testing PCA and GPR to remove the residual foregrounds, we find that if bright sources with flux density $>10\,$mJy are subtracted with the rest of the sources being modelled to 90\% accuracy, removing the residual foreground can enable detections of the \hi\ power spectrum at large scales. The foreground wedge is consistent with the instrinsic foreground power coupled with the instrument chromaticity, with the wedge corresponding to the primary beam size.
    \item
    Assuming no contribution from thermal noise and systematic effects, the empirical data covariance matrix calculated from the image cube reflects the true underlying covariance of the sky signal. Therefore, PCA and GPR can both sufficiently remove the foregrounds with trivial differences between these two methods.
    \item
    From the image cube with $\rm (1.5\,deg)^2$ sky area within the primary beam FoV, we can measure the \hi\ power spectrum from $k\sim 0.02\,{\rm Mpc^{-1}}$ to $k\sim 0.3\,{\rm Mpc^{-1}}$.
\end{itemize}

The results suggest that measuring the \hi\ power spectrum at $5<z<6$ for cosmological analysis using SKA-Low is viable and will open up a new window for cosmology in the near future. Using wide-field imaging and/or mosaicing, we can probe linear cosmological scales $k\sim 0.01\,{\rm Mpc^{-1}}$ to quasi-linear scales $k\sim 0.3\,{\rm Mpc^{-1}}$. The wide range of clustering scales probed can be used to constrain cosmology \citep{2023MNRAS.519.6246P}. 

The detection of the \hi\ signal at large cosmological scales depends heavily on the robustness of foreground mitigation strategies. Simulating different level of depths for the observation, we find that:
\begin{itemize}
    \item 
    In general, future observations using SKA-Low contain a high level of thermal noise fluctuations. The effects of the thermal noise on the data covariance are visible even for deep observations $>250$ h.
    \item
    The thermal noise fluctuations in the empirical data covariance matrix make residual foreground removal more difficult. Thermal noise creates numerical features on the foreground-removed image cube on small frequency scales, breaking the spectral smoothness of the data covariance.
    \item
    As a result of the spectral fluctuations, foreground removal methods induce numerical artefacts on small frequency scales. The numerical artefacts leak power into the observation window which leads to significant bias on the \hi\ power spectrum estimation. Even scales $k_\parallel >0.1\,{\rm Mpc^{-1}}$ which can be probed with just foreground avoidance can be contaminated.
    \item
    Comparing PCA and GPR, we find that GPR performs much better in the presence of thermal noise. The key factor is that GPR uses smooth kernels to model the signal and apply the fitted kernels instead of the actual data covariance matrix for the foreground removal. For observation with integration time $>250$ h, GPR can sufficiently remove the foregrounds and allow unbiased estimation of the \hi\ power spectrum for $k_\parallel>0.3k_\perp$ regions.
    \item
    For the integration time of 600 h, SKA-Low will be able to measure the \hi\ power spectrum in the $5<z<6$ bin from $0.03$ to $0.3\,{\rm Mpc^{-1}}$ with a signal-to-noise ratio of $\sim 5$ across the scales.
\end{itemize}

In conclusion, the viability of detecting the cosmological \hi\ power spectrum at $5<z<6$ using SKA-Low depends on deep observations to preserve the spectral smoothness of the data covariance to facilitate sufficient foreground removal. It will allow accurate measurement of the \hi\ power spectrum, on the premise that deep fields with effective integration time $\gtrsim 300$ h are observed. Our results not only solidify the science case of measuring post-reionization cosmology with SKA-Low, but also provides insights into survey design for maximising the scientific output of the instrument.

Finally, we provide a qualitative study into the systematic effects by introducing spectral fluctuations that can originate from bandpass instabilities and calibration errors. Testing the data analysis pipeline for different levels of systematics we find that:
\begin{itemize}
    \item
    Systematic effects such as bandpass instabilities will introduce fluctuations in the small frequency interval, breaking the spectral smoothness of the foregrounds. It leads to spillover of the foreground power into the observation window outside the foreground wedge.
    \item
    In the image cube averaged across all timesteps, the effective systematic errors acorss the frequency channels need to be small to suppress the contamination. If the level of the systematics is above $10^{-4}$, the power spectrum measurement will be biased across all scales of interests.
    \item
    For systematic errors $\lesssim 5\times 10^{-5}$, we find that using GPR to perform foreground removal gives unbiased estimation of the \hi\ power spectrum.
\end{itemize}
The requirements on containing the systematic errors below one per cent level again highlight the need for deep observations with good understanding of the sky model and the instrument. With the unprecedented power of the SKA-Low array, we expect that future surveys will be sufficiently systematic-mitigated to enable the detection of the \hi\ power spectrum for the high redshift, post-reionization Universe.

Our work strongly favours using the future SKA-Low data for \hi\ science after cosmic reionization. We have demonstrated that the \hi\ power spectrum can be measured with statistical significance using observational depth that can easily be reached using SKA-Low. Furthermore, we have showcased residual foreground removal using GPR that suppresses the foreground wedge to probe cosmological scales, which is robust in the presence of a reasonable level of systematic effects. The tools presented in this paper can be further used for more realistic simulations of SKA-Low observations to develop the data analysis pipeline towards future detections.

\section*{Acknowledgements}
We thank Keith Grainge and Mike Wilensky for discussions. EC acknowledges the support of a Royal Society Dorothy Hodgkin Fellowship and a Royal Society Enhancement Award. LW is a UK Research and Innovation Future Leaders Fellow [grant MR/V026437/1]. AM acknowledges the support of a UK Research and Innovation Future Leaders Fellowship [grant MR/V026437/1]. Apart from aforementioned packages, this work also uses \textsc{pytorch} \citep{NEURIPS2019_9015}, \textsc{numpy} \citep{2020Natur.585..357H}, \textsc{scipy} \citep{2020NatMe..17..261V}, \textsc{astropy} \citep{2018AJ....156..123A}, \textsc{camb} \citep{2000ApJ...538..473L}, \textsc{emcee} \citep{2013PASP..125..306F} and \textsc{matplotlib} \citep{Hunter:2007}. 

\section*{Data Availability}
Data underlying this paper will be shared on reasonable request to the corresponding author.



\bibliographystyle{mnras}
\bibliography{example} 




\appendix

\section{Quadratic Estimator for Power Spectrum Estimation}
\label{sec:est}
We present the quadratic estimator for the power spectrum estimation used in this paper, following Eq. (\ref{eq:tf}) and (\ref{eq:pscal}). The aim of using the quadratic estimator formalism is to incorporate renormalisation of the estimator after the operations of foreground cleaning and frequency tapering. It also performs bias correction to remove the thermal noise power spectrum and potentially some bias from the GPR cleaning. Our formalism follows closely the work of \cite{2021MNRAS.501.1463K} and \cite{2023MNRAS.518.2971C}. Note that, we are not aiming to construct the covariance for the total data vector with the number of elements being the number of pixels times the number of frequency channels $N_{\rm pix}\times N_{\rm f}$. The resulting covariance matrix of size $(N_{\rm pix}\times N_{\rm f})^2$ is too large and therefore not of our interests for a preliminary study. Instead, we construct the estimator for each pixel across the $k_\parallel$ direction, so that we are only dealing with one pixel at a time with a covariance matrix of size $N_{\rm f} \times N_{\rm f}$.

In this section, we use $i$ to denote the $i^{\rm th}$ pixel in the Fourier transformed image cube. For each pixel, the Fourier density gives a bandpower vector $(\hat{\mathbf{p}}^{i}_{\rm T})_{\alpha}$, with the $\alpha^{\rm th}$ element being the power spectrum at $({\bm{k}}_\perp^i,k_\parallel^{\alpha})$. The quadratic estimator can be written as
\begin{equation}
    (\hat{\mathbf{p}}^{i}_{\rm T})_{\alpha} =\big(\tilde{\mathbf{d}}^{i}\big)^{\dagger}\mathbf{E}_{\alpha}^{i}\tilde{\mathbf{d}}^{i}-\hat{\mathbf{b}}_{\alpha},
\label{eq:est}
\end{equation}
where $\mathbf{E}_{\alpha}^{i}$ and $\hat{\mathbf{b}}_{\alpha}$ are the estimation matrix and bias correction respectively. Here, $\tilde{\mathbf{d}}^{i}$ is the data vector along the frequency direction for the $i^{\rm th}$ pixel. We collapse the Fourier transform along the transverse directions and the PSF deconvolution in this data vector so that for the $j^{\rm th}$ frequency channel
\begin{equation}
\begin{split}
    (\tilde{\mathbf{d}}^{i})_{j} = &\int \frac{{\rm d}^2 x_\perp}{{\rm \mathbf{V}}} {\rm exp}\big[-i\bm{k}_\perp^i \cdot \bm{x}_\perp \big]\bigg(\frac{\lambda^2}{\rm 2k_B}\bigg)^2 \frac{I(\bm{x}_\perp,x_\parallel^j)}{A^2(\bm{x}_\perp,x_\parallel^j)} \\ & \bigg/ \widetilde{\rm PSF}(\bm{k}_\perp^i,f_{\rm c}).
\end{split}
\label{eq:2dfourier}
\end{equation}

The estimation matrix $\mathbf{E}_{\alpha}^{i}$ can be written as
\begin{equation}
    (\mathbf{E}_{\alpha}^{i})_{\alpha} = \sum_{\beta} \mathbf{M}_{\alpha \beta} \mathbf{R}^\dagger \mathbf{T}^\dagger \mathbf{F}^\dagger w_\beta \mathbf{F}\, \mathbf{T}\,\mathbf{R} = \sum_\beta \mathbf{M}_{\alpha \beta} \mathbf{R}^\dagger \mathbf{T}^\dagger \mathbf{C}_{,\beta} \mathbf{T} \mathbf{R},
\end{equation}
where $\mathbf{M}_{\alpha \beta}$ is the renormalisation matrix, $\mathbf{T}$ is the frequency taper, $w_\beta$ is the selection matrix with all elements being zero except the $\beta^{\rm th}$ diagonal element and $\mathbf{F}$ is the 1D discrete Fourier transform kernel along the frequency direction. $\mathbf{C}_{,\beta} = \mathbf{F}^\dagger w_\beta \mathbf{F}$ is the Fourier operator. $\mathbf{R}$ is the foreground removal operation. For PCA as described in Eq. (\ref{eq:pca}), the removal matrix is $\mathbf{R} = \mathbf{I} - \mathbf{A}\mathbf{A}^{\rm T}$ where $\mathbf{I}$ is the identity matrix. For GPR as described in Eq. (\ref{eq:gpr}), the removal matrix is  $\mathbf{R} = \mathbf{I} - {\mathbf{K}}_{\rm fg}\big(\mathbf{K}_{\rm fg}+\mathbf{K}_{\rm n}+\mathbf{K}_{\rm \hi}\big)^{-1}$.

The renormalisation matrix can be calculated by taking the expectation value of Eq. (\ref{eq:est})
\begin{equation}
    \langle (\hat{\mathbf{p}}^{i}_{\rm T})_{\alpha} \rangle = \sum_\beta {\rm tr}\big[ \mathbf{C}_{, \beta} \mathbf{E}_\alpha^i \big](\mathbf{p}^i_{\rm T})_\beta + {\rm tr}\big[\big(\mathbf{N} + \mathbf{C}_{\rm fg}\big) \mathbf{E}^{\rm d}_\alpha \big] - \hat{b}^{\rm d}_\alpha.
\label{eq:estwindow}
\end{equation}
Following \cite{2021MNRAS.501.1463K}, we can form the quantity
\begin{equation}
    H_{\alpha\beta} = {\rm tr}\big[ \mathbf{R}^\dagger \mathbf{T}^\dagger \mathbf{C}_{,\alpha} \mathbf{T}\, \mathbf{R} \mathbf{C}_{,\beta}\big],
\end{equation}
and choose $\mathbf{M} = \mathbf{H}^{-1/2}$ \citep{2002MNRAS.335..887T} to renormalise the estimator.

In order to remove the bias in the power spectrum estimation from the foregrounds and the thermal noise, from Eq. (\ref{eq:estwindow}) we can choose
\begin{equation}
    \hat{b}^{\rm d}_\alpha = {\rm tr}\big[\big(\mathbf{N} + \mathbf{C}_{\rm fg}\big) \mathbf{E}^{\rm d}_\alpha \big]
\end{equation}
to remove the bias. In reality though, we do not know the underlying thermal noise and the foregrounds. In order to remove the noise bias, we calculate $\mathbf{N}$ by simulating 1000 realisations of the thermal noise using the same $\sigma_{\rm N}$. Here, $\sigma_{\rm N}$ is assumed to be a known quantity, which is the case for our simulation. In real observations, a good estimation of $\sigma_{\rm N}$ can be obtained by calculating the fluctuations of the Stokes V visibility data on long baselines (e.g. \citealt{2021MNRAS.505.2039P}). For each realisation, we pass the visibility data to the same imaging pipeline to generate the image cubes. For each pixel in the image cube, we then calculate the average frequency-frequency correlation across all realisations to obtain an estimation of $\mathbf{N}$. We bin the resulting noise bias ${\rm tr}[\mathbf{N} \mathbf{E}^{\rm d}_\alpha ]$ into cylindrical $k$-space and show the thermal noise power spectrum in Fig. \ref{fig:bias} for the case with 360 h of integration time. The vertical stripes follow the baseline densities on different scales.

\begin{figure}
    \centering
	\includegraphics[width=\columnwidth]{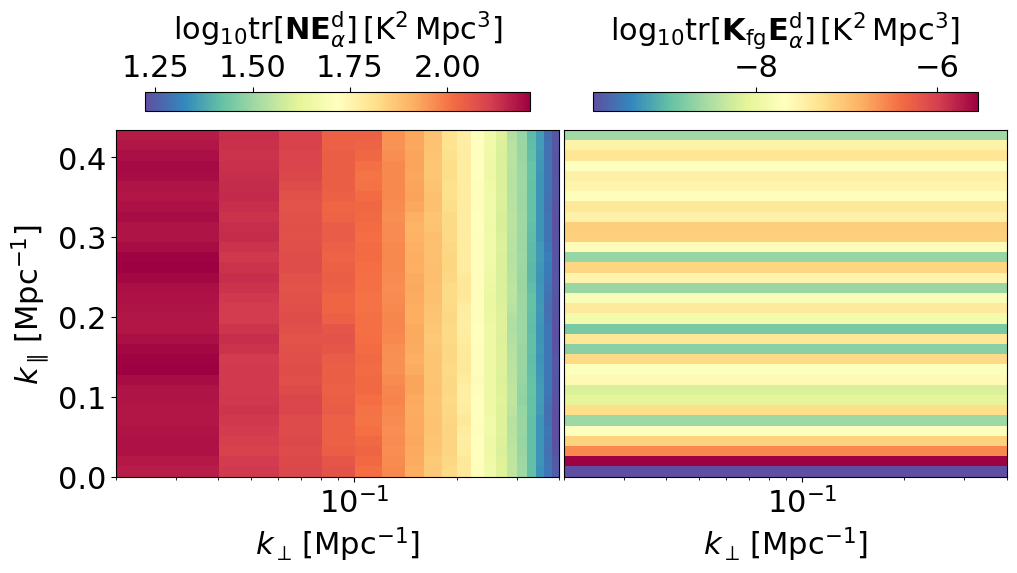}
    \caption{Left panel: The cylindrical power spectrum of the thermal noise calculated using ${\rm tr}[\mathbf{N} \mathbf{E}^{\rm d}_\alpha ]$ for the case of 360 h of integration time. Right panel: The cylindrical power spectrum of the GPR bias correction using ${\rm tr}[\mathbf{K}_{\rm fg} \mathbf{E}^{\rm d}_\alpha ]$ for the case of 360 h of integration time.  }
    \label{fig:bias}
\end{figure}

The covariance of the foregrounds can be extracted from the GPR fitted kernel $\mathbf{K}_{\rm fg}$. Note that, since we work on the frequency-frequency covariance, $\mathbf{K}_{\rm fg} \mathbf{E}^{\rm d}_\alpha$ is the same for each pixel, and therefore there is no $k_\perp$ dependence of the bias term as shown in the right panel of Fig. \ref{fig:bias}. We note that this is not a result of GPR but the result of our simplified quadratic estimator formalism. Nevertheless, it gives us a good estimation of the order of magnitude of the GPR bias correction. As one can see, the correction is at least 2 orders of magnitude smaller than the \hi\ signal shown in Fig. \ref{fig:hi30n}, and therefore this bias correction is negligible in our case.

Finally, we comment on the fact that in the power spectrum estimation, the 2-D Fourier transform shown in Eq. (\ref{eq:2dfourier}) is applied to the data before the GPR removal $\mathbf{R}$, while the GPR fitting for the kernels are done on the original data vector before the transform. These two operations are commutable, as the GPR removal only operates along the frequency direction, independent of the 2-D Fourier transform on the transverse plane. We verified that there is no visible difference in the resulting power spectrum if these two operations are swapped. Performing the 2-D Fourier transform first allows us to only construct the estimator one pixel at a time, providing massive speed-up.

\section{Caveats of the simulations}
\label{sec:caveat}

We discuss the limitations of our simulation settings. Specifically, we quantify the effects of limited $(10.5\,{\rm deg})^2$ sky area for the input signal, coupled with the instrument beam which gets cut off at 1\% at the $10.5\,{\rm deg}$ angular extent. Furthermore, we discuss the Gaussian calibration errors simulated in terms of its structure in frequency.

The primary beam of the instrument is shown in Figure \ref{fig:beam_large}. Around the centre $(10.5\,{\rm deg})^2$ region, the beam only goes down to $10^{-2}$, introducing sharp features in the simulation. We first note that, as discussed in Section \ref{sec:fgwedge}, the image power spectrum does not show a clear wedge structure due to the small image size. To investigate the chromatic structure of the data, we instead calculate the delay power spectrum directly from visibility and present it in Figure \ref{fig:delay}. As shown in the top panel, the full foreground delay power spectrum shows a clear wedge structure. Above the wedge, the effect of sky signal getting cut off at $10.5\,{\rm deg}$ can be seen as the diamond-shape structures. Assuming the bright sources are removed as described in Section \ref{sec:skysim}, we calculate the delay power spectrum of the residual foregrounds shown in the centre panel of Figure \ref{fig:delay}. The chromatic features disappear as there is no bright emission coming from the beam sidelobes. Finally, we also present the delay power spectrum of the \hi\ signal in Figure \ref{fig:delay} to show that the sky cut-off does not affect the \hi\ simulation. 

\begin{figure}
    \centering
	\includegraphics[width=\columnwidth]{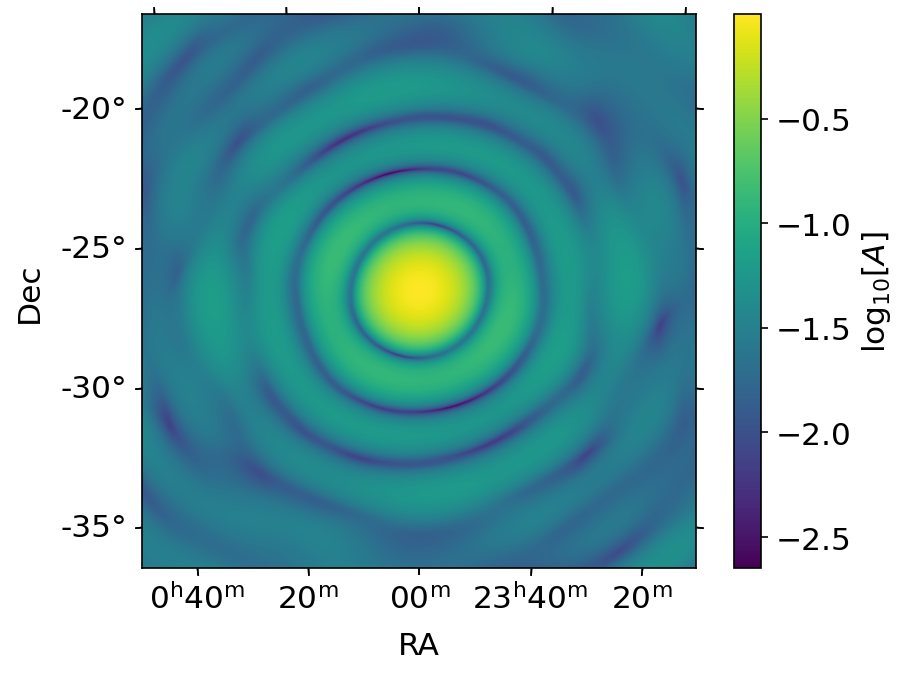}
    \caption{The primary beam of the instrument around the pointing centre in our simulation. The image size is $\rm (20\, deg)^2$. The beam is simulated at the central frequency 220MHz and averaged over all time steps for one station.}
    \label{fig:beam_large}
\end{figure}

\begin{figure}
    \centering
	\includegraphics[width=\columnwidth]{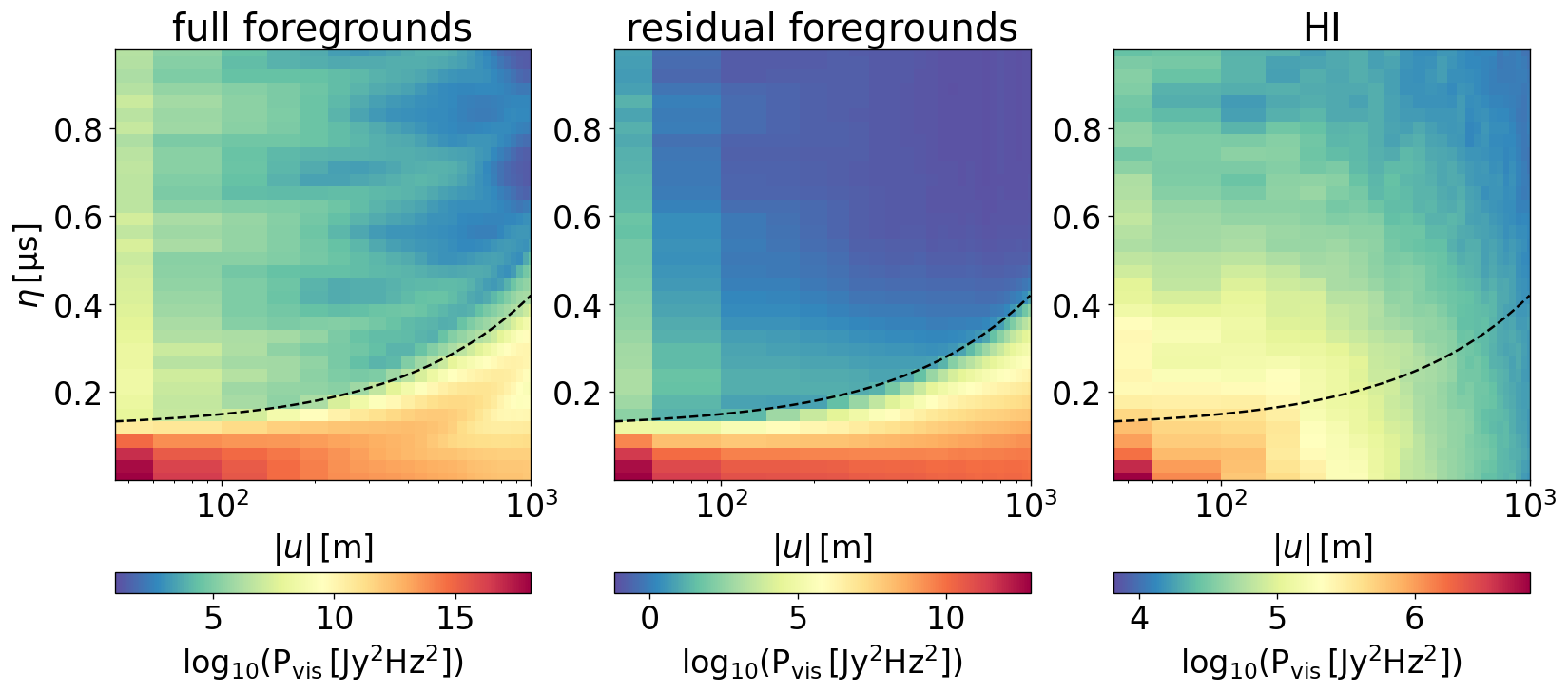}
    \caption{The delay power spectra of the visibility data in our simulation. Left panel: The delay power spectrum of the full foreground signal. Central panel: The residual foregrounds. Right panel: the \hi\ signal. The $|u|$ and $\eta$ range correspond roughly to the $k$-range of the cylindrical power spectra shown in the paper. The black dashed line denotes the foreground wedge.}
    \label{fig:delay}
\end{figure}

The calibration errors for SKA-Low observations are likely smooth in frequency \citep{2019ApJ...875...70B}, which are not the Gaussian random fluctuations we use. Using the delay power spectra, we then investigate the assumption of the Gaussian gain errors described in Section \ref{sec:vissim}. For comparison, we simulate another type of error that follows the sine function with a period of 15 frequency channels. The errors are then rescaled so that the standard deviation across the channels is $10^{-4}$. The sine errors are then compared against the Gaussian errors as shown in Figure \ref{fig:scatter}. 

\begin{figure}
    \centering
	\includegraphics[width=\columnwidth]{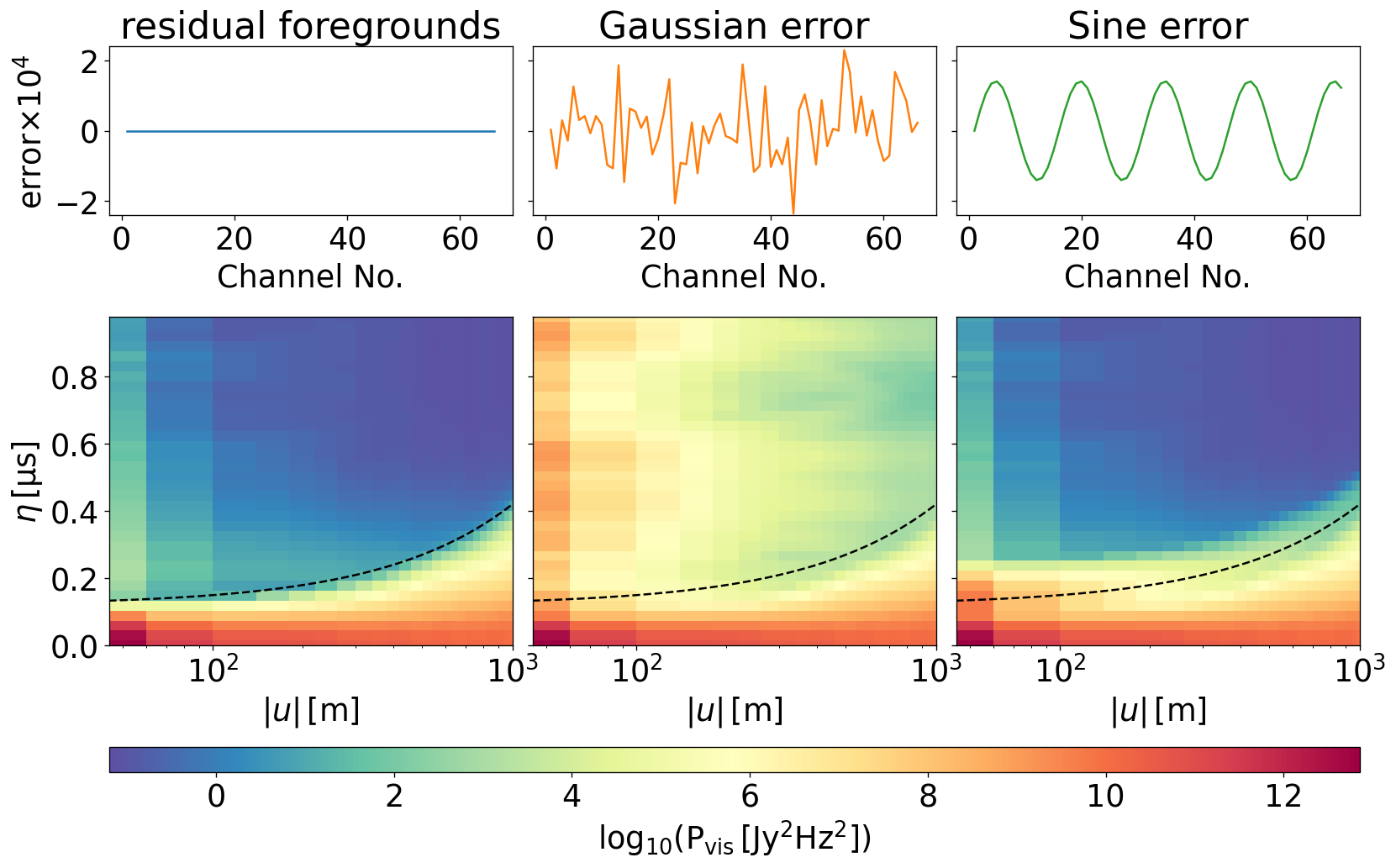}
    \caption{The delay power spectra of the visibility data after applying gain errors. Left panel: The delay power spectrum of the residual foreground signal with no error applied. Central panel: The residual foregrounds with Gaussian errors as shown in the top and the delay power spectrum shown in the bottom. Right panel: The residual foregrounds with sine errors.}
    \label{fig:scatter}
\end{figure}

For the sine error case shown in the right panel, the foreground wedge gets lifted into higher delay. Comparing to the Gaussian error case in the central panel, the leakage still concentrates around relatively low delay. This means that the foreground contamination can be easier to remove for GPR, as its structure has large frequency intervals. In the Gaussian case, however, the scatter of the foreground power into higher delay is visible across all scales. The foreground contamination is at the smallest frequency interval, which is difficult to distinguish from the \hi\ signal. Therefore, we conclude that the conclusions reached in Section \ref{subsec:system} are robust, as the foreground contamination is not an optimistic case.

We emphasize that, the smooth frequency structures of the gain errors pose other challenges in sky modelling and continuum subtraction, which are beyond the scope of this paper and left for future work.

\bsp	
\label{lastpage}
\end{document}